\documentclass{aastex62}

\usepackage{natbib}
\usepackage{amsbsy}
\usepackage{amssymb}
\usepackage{xcolor}
\usepackage{comment}
\received{February 6, 2019}
\revised{March 9, 2019}
\accepted{March 19. 2019}
\submitjournal{ApJ}

\shorttitle{Claims Concerning Solar Rotation}
\shortauthors{Scherrer and Gough}

\begin{document}
\title{A Critical Evaluation of Recent Claims Concerning Solar Rotation}

\correspondingauthor{Philip Scherrer}
\email{pscherrer@solar.stanford.edu,douglas@ast.cam.ac.uk}

\author[0000-0002-6937-6968]{P.H. Scherrer}
\affil{Physics Department, Stanford University \\
452 Lomita Mall, Stanford, CA 94305-4085 USA} 

\author{D.O. Gough}
\affiliation{Institute of Astronomy, and Department of Applied Mathematics and Theoretical Physics, University of Cambridge \\
Madingley Road, Cambridge, CB3 0HA, UK}
\affil{Physics Department, Stanford University \\
452 Lomita Mall, Stanford, CA 94305-4085 USA}



\begin{abstract}
\citet{fossat_etal_2017} recently reported detecting 
rotational splitting of g modes indirectly via the interaction with 
p modes observed directly by the GOLF instrument on SOHO. 
They concluded that the core of the Sun is rotating $3.8 \pm 0.1$ 
times faster than the surrounding radiative envelope. 
This is startling, partly because such rapid rotation almost contradicts direct inferences from the p-mode rotational 
splitting inferred from the same data. Moreover, the inferred amplitudes of the g modes appear to
exceed the upper bound reported by \citet{appourchauxetal_gmode_limit_2010A&ARv..18..197A}. It is also suspect because the theory of the procedure implies that the principal modes claimed to have been measured should be 
undetectable. We point out that there are other interpretations: one
leads to a core rotation about twice faster than the surrounding envelope; another, to a core rotating more slowly than 
the envelope.
 Here we also report on an independent assessment of the Fossat et al. analysis
by applying their procedure to different representations of the GOLF data, expanding on \citet{schunker_fragile_g-mode_detection2018SoPh..293...95S}.
We also analyze seismic data 
obtained from LOI and MDI (both also on SOHO), from HMI (on SDO) and from the ground-based BiSON and GONG, and found the evidence reported by Fossat et al. not to be robust. We also illustrate that merely fitting model spectra to observations, 
which Fossat et al. do to support their g-mode detections and as 
\citet{fossatschmider2018A&A} do for extracting additional g-mode 
splittings, is not necessarily reliable. We are therefore led to doubt the claim.

\end{abstract}



\section{Introduction}
\label{sec:intro}

The density and pressure perturbations, and the associated velocity
field, produced by g modes modify the propagation of acoustic waves and so influence the p-mode frequencies.  As \citet{kennedy_etal_1993} have pointed out, that, in principle, offers a potential probe of the dynamics of solar g modes of even degree, and may be the only way in which accurate 
constraints on the structure and kinematics of the innermost regions of the Sun can
be accessed.  

The influence of high-order g modes of low degree can most easily be estimated by 
noting that the g-mode frequencies are much lower than those of the p modes, and that therefore 
to a first approximation the effectively instantaneous p-mode frequencies
$\omega$ 
can be estimated by ignoring the explicit g-mode time dependence.   
The time dependence of the global angular velocity can be ignored too.  
In that extreme case, g modes modify 
the frequency of any particular p mode independently of the others,  
and then the variational integral 
relation of \citet{dlbjpo1967MNRAS} can be brought to bear to characterize the 
p-mode frequencies: 
\begin{equation}
{\cal I}\omega^2-2{\cal R}\omega-{\cal K}=0\, ,
\label{1.1}
\end{equation}
where ${\cal I}({\pmb \xi^*,\pmb \xi})$, ${\cal R}(\pmb \xi^*,\pmb \xi)$  and ${\cal K}(\pmb \xi^*,\pmb \xi)$ are integrals 
depending on the instantaneous underlying state of the star, including the (small) contribution from the g modes but 
not from the p modes. They depend explicitly on the (complex) adiabatic p-mode displacement eigenfunction 
$\pmb\xi($\textbf{\textit{r}}$){\rm e}^{{\rm -i}\omega t}$, where $\textbf{\textit{r}}$ is a position vector in an inertial frame centred on the star,  and $t$ is time. Also 
${\rm i} = \sqrt(-1)$. The integrals are over the volume $\cal V$ of the star;  
they are given approximately by
\begin{equation}
{\cal I}({\pmb \xi^*,\pmb \xi}) = \int_{\cal V} \pmb \xi^*.\,\pmb \xi \rho {\rm d}V\, ,
\label{1.2}
\end{equation}
\begin{equation}
{\cal R}({\pmb \xi^*,\pmb \xi}) = {\rm i} \int_{\cal V} \pmb \xi^*.(\textbf{\textit{v}}.\nabla \pmb \xi) \rho {\rm d}V
\label{1.3}
\end{equation}
and
\\
\begin{equation}
{\cal K}({\pmb \xi^*,\pmb \xi}) = \int_{\cal V} [\rho c^2 {\rm div}{\pmb \xi^*} {\rm div}\pmb \xi + (\pmb \xi^*   {\rm div}{\pmb \xi} + \pmb \xi {\rm div}{\pmb \xi^*}). \nabla p  + \rho^{-1} \pmb \xi^* .\nabla \rho \; \pmb \xi . \nabla p ]\,{\rm d}V
\label{1.4}
\end{equation}
\citep[cf.][]{dog_L-H1993}, in which $p$ is pressure, $\rho$ is density and $c$ is the adiabatic sound speed.  The asterisk denotes complex conjugate.  The vector 
$\textbf{\textit{v}}$ is the background fluid velocity, which accommodates both the angular velocity $\Omega(r,\theta)$, with respect to spherical polar coordinates $(r,\theta,\phi)$, and the (real) velocity $\boldsymbol{u}$ associated with the g modes. For simplicity, we ignore steady meridional circulation.   
We are therefore presuming the angular velocity to be about the unique axis about which the 
co-ordinates $(r, \theta, \phi)$ are defined. 
Because both the p modes and the g modes are of high order, 
the Cowling approximation (in which the Eulerian perturbation to the gravitational potential is ignored) has been adopted, and for added simplicity a small surface boundary 
term has been neglected.  It would be straightforward to include them. In addition, 
terms quadratic in $\Omega(r,\theta)$ and in the 
g-mode perturbation are also neglected.  
As \citet{dlbjpo1967MNRAS} demonstrated, the integral 
$\cal R$ is  skew symmetric, and $\cal I$ and $\cal K$ are symmetric in $\xi^*$  and $\xi$;  
equation (\ref{1.1}) constitutes a variational principle for the p modes,  and therefore it is 
sufficient to employ the displacement eigenfunction $\pmb \xi$ of what we call the basic state, namely a 
spherically symmetric model of the Sun (in which $\nabla p$ and $\nabla \rho$ are both radially directed,  and ${\cal K}$ 
is therefore obviously symmetric), unperturbed by rotation and the modes of oscillation.

  The eigenfunctions of the basic state are of the form
\begin{equation}
\pmb \xi_{n,l,m}({\boldsymbol{r}})=\left(\xi_{n,l}(r)P^m_l,\frac{\eta_{n,l}(r)}{L}\frac{{\rm d}P^m_l}{{\rm d}\theta},\frac{{\rm i}m\eta_{n,l}(r)}{L\, {\rm sin}\theta}P^m_l\right){\rm e}^{{\rm i}m\phi}\,,
\label{1.6}
\end{equation}
\begin{equation}
\psi_{n,l,m}({\boldsymbol{r}})=\psi_{n,l}(r)P^m_l{\rm e}^{{\rm i}m\phi}\,,
\label{1.7}
\end{equation}
where $\psi$ is a scalar component of the oscillation eigenfunction such as the pressure or density perturbation; 
$P^m_l({\rm cos}\theta)$ is the associated Legendre function of the first kind, 
of degree $l$ and order $m$; and $L=\sqrt{l(l+1)}$.  We adopt the normalization 
$P^{-m}_l({\rm cos}\theta)=P^m_l({\rm cos}\theta)$. 
The modes can be characterized by 
their order $n$, degree $l$ and azimuthal order $m$.  We note that the modes are 
degenerate in $m$: any group of modes of like $n$ and $l$ and varying $m$, having 
a common frequency, is called a multiplet; the individual modes are singlets. Degeneracy 
is lifted by a dynamically pertinent aspherical perturbation to the basic state.

The multiplet cyclic frequencies $\nu _{n,l}$ of acoustic modes (p modes) of low degree and asymptotically high order have cyclic eigenfrequencies $\nu\, (= \omega/2\pi)$  
given by
\begin{equation}
\nu_{n,l} \sim \left(n+\frac{1}{2} l+\alpha_{\rm p}\right)\nu_0-(A_{\rm p}L^2-B_{\rm p})\frac{\nu_0^2}{\nu_{n,l}} + ...\,,
\label{1.7a}
\end{equation}
where
\begin{equation}
\nu_0^{-1}=2\int_0^Rc^{-1}{\rm d}r
\label{1.7b}
\end{equation}
is the sound travel time across a solar diameter, 
and  $\alpha_{\rm p}$, $A_{\rm p}$ and $B_{\rm p}$ are dimensionless constants of order unity depending
only on the basic state;  
a  g mode, whose order $n$  is formally negative (although we shall loosely refer to $|n|$ as the g-mode order), of low-degree $l$ with asymptotically high $|n|$ has period 
$P_{n,l}$, also degenerate in azimuthal order $m$, given by 
\begin{equation}
LP_{n,l} \sim \left(|n|+\frac{1}{2} l+\alpha_{\rm g}\right)P_0-(A_{\rm g}L^2-B_{\rm g})\frac{P_0^2}{P_{n,l}} + ...\,,
\label{1.7c}
\end{equation}
in which  
\begin{equation}
P_0=2\pi^2\left(\int_0^R \frac{N}{r} {\rm d}r\right)^{-1}\,,
\label{1.7d}
\end{equation}
$N=\sqrt{g(H^{-1}-g/c^2)}$ being the buoyancy (Brunt-V\"ais\"ala) frequency;  
$\alpha_{\rm g}$, $A_{\rm g}$ and $B_{\rm g}$ 
are also constants of order unity depending only on the basic state \citep{tassoulasymptotics1980ApJS.43.469T,provostberthomieu1986A&A...165..218P,ellis_g-modes_1988IAUS..123..147E,dog_L-H1993}. Here, $g$ is the local 
acceleration due to gravity and $H$ is the density scale height.  The coefficients $A_{\rm p}$ and $A_{\rm g}$, multiplying $L^2$, depend in a rather simple way on 
the stratification predominantly near the center of the Sun; the remaining coefficients depend on the 
stratification principally near the pertinent upper turning points, near the surface of the Sun in the case of $\alpha_{\rm p}$ and $B_{\rm p}$, and in the vicinity of the base of the convection zone in the case of $\alpha_{\rm g}$ and $B_{\rm g}$.  
Thus, $\alpha_{\rm p}$ and $B_{\rm p}$ depend, for example on the uncertain turbulent
boundary layer at the top of the convection zone, and 
$\alpha_{\rm g}$ and $B_{\rm g}$ on the details of the mixing
process in the tachocline. Some typical values are discussed in the aforementioned
references.
The GOLF (Global Oscillations at Low Frequency, \citet{Gabriel1995})  instrument on SOHO (Solar and Heliospheric Observatory, \citet{1995SoPh..162....1D})   is sensitive primarily to p modes of degrees $l \le 3$ with little sensitivity for degrees $l = 4,5$.
Thus, an incompletely resolved power spectrum is a sequence of peaks, composed  alternately of even- and odd-degree components, almost uniformly spaced in frequency by approximately $\nu_0/2$.  In contrast, g modes form sequences approximately uniformly 
separated in period for each value of $l$.

It is straightforward to estimate the g-mode-induced perturbations, $\delta_{\rm g} \omega$, to the instantaneous p-mode frequencies by  perturbing equation (\ref{1.1}) about the basic state, and retaining only terms linear in the perturbation.  The outcome is
\begin{equation}
\delta_{\rm g}  \omega = [{\cal R}_{\boldsymbol{u}}+(\delta{\cal K}-\omega^2\delta{\cal I})/2\omega]/{\cal I}\,
\label{1.5}
\end{equation}
in which ${\cal R}_{\boldsymbol{u}}$ is the component of ${\cal R}$ arising solely from 
the g-mode velocity eigenfunction of the basic state; the structure perturbations 
$\delta{\cal I}$ 
and $\delta{\cal K}$ are also computed from the g-mode eigenfunctions.  It is presumed 
that such perturbations, determined from the GOLF data, are what have been analyzed by \citet{fossat_etal_2017}
to obtain the rotational splitting of the g-mode frequencies.

In preparation for our discussion of the analysis, we record first the frequency splitting arising directly from the Sun's rotation. It is obtained by setting 
${\boldsymbol{v}}=\Omega{\boldsymbol{k}}\times{\pmb r}$ in equation (\ref{1.3}), where ${\boldsymbol{k}}$ is a unit vector parallel to what we presume to be the unique rotation axis.  The outcome is 
\begin{equation}
\delta_\Omega \omega = {\cal R}_\Omega/{\cal I},
\label{1.8}
\end{equation}
where 
\begin{equation}
{\cal R}_\Omega= \int_{\cal V} [m{\pmb \xi^*}.{\pmb \xi}+{\rm i}{\boldsymbol{k}}.({\pmb \xi^*}\times{\pmb \xi})]\Omega\rho{\rm d}V\,.
\label{1.9}
\end{equation}

For p modes of low degree and 
high order, which are the diagnosing modes of interest here, the displacement eigenfunctions 
are predominantly vertical nearly everywhere; therefore their vector products are nearly zero, rendering the 
second term in the integrand in equation (\ref{1.9}) much smaller than the first.  
Consequently, the p-mode rotational 
splitting per unit azimuthal order is approximately the eigenmode-energy weighted volume average $<\Omega>_{\rm p}$ 
of the global angular velocity $\Omega$:
\begin{equation}
\frac{\delta_{\Omega {\rm p}}\omega}{m} \simeq {\cal I}^{-1}\int_{\cal V} {\pmb \xi^*}.{\pmb \xi}\Omega\rho{\rm d}V\,=: \,<\Omega>_{\rm p}\,.
\label{1.10}
\end{equation}
Were the latitudinal variation of 
$\Omega$ to have been neglected, that average would have been approximately a radial average weighted by the acoustic slowness (i.e. inverse sound speed) when $n/L \gg 1$;  in 
that limit,  $<\Omega>_{\rm p}$ is independent of the degree, azimuthal order and frequency of the p mode 
in question. 
\citet{fossat_etal_2017} presumed instead its value to be that which, when multiplied by the moment of inertia $I$ of the Sun, yields the total angular momentum $H$: 
\begin{equation}
<\Omega>_{\rm p,Fossat et al.}=H/I,
\label{1.10a}
\end{equation}
which does not weight the angular velocity $\Omega$ in the same way as do the p-mode inertial terms inducing rotational splitting, 
although numerically that matters little if the only region where $\Omega$ deviates substantially from its near uniform value in the 
radiative envelope is in a small central core. 
The actual contributions from the core, defined here as the region in which most of the nuclear energy is generated, to both ${\cal I}$ and $I$ are fairly 
small (between about $4\%$ and $7\%$ to $\cal I$ from the p-modes considered here, and about $6\%$ to $I$, if the core radius $r_{\rm c}$ is taken to be $20\%$ of the radius $R$ 
of the Sun).

For g modes whose order is much greater in 
magnitude than their degree, the horizontal component of the displacement eigenfunction  is dominant.   It follows 
immediately from equations (\ref{1.10}) and (\ref{1.6}) that if
the latitudinal variation of $\Omega$ can be ignored, the frequency splitting is given 
approximately by
\begin{equation}
\frac{\delta_{\Omega {\rm g}}\omega}{m} \simeq (1-L^{-2})<\Omega>_{\rm g}\,,
\label{1.11}
\end{equation}
where $<\Omega>_{\rm g}$ 
is again the eigenmode-energy weighted volume average of $\Omega$, this time appropriate to g modes; when the magnitude of 
$n/L$ is large, it is approximately a radial average (in the radiative interior) weighted by $N/r$, 
which is independent of the g-mode frequency.  
The error in equation (\ref{1.11}) resulting from the latitudinal variation of the 
angular velocity in the convection zone is small, because the g modes of interest 
are evanescent there.  
Therefore the g-mode frequency spectrum should contain combs of approximately uniformly separated frequencies as  $m$ varies, spaced differently for different $l$.

In the analysis that follows, we shall assume that $\Omega$ does not vary with latitude in the radiative interior.  
It is a good first approximation, and is consistent with the assumptions of \citet{fossat_etal_2017}.

\section{The procedure of Fossat et al., and a comment on their inference}
\label{sec:procedure}
Broadly speaking, the p-mode frequencies are most sensitive to conditions in the 
upper layers of the convection zone (yet beneath the upper turning points), where the sound speed is lowest and the energy density the greatest. They are therefore susceptible 
to temporal variations in the stratification of the outermost regions of the Sun caused by 
solar activity,  which interferes with attempts 
to detect g modes in the Sun's radiative interior.  However, p modes with similar frequencies would 
likely be influenced in a similar manner, at least were the Sun to be spherically symmetrical. So \citet{fossat_etal_2017} investigated the 
time dependence of the so-called large frequency separation\footnote{Actually, half the large separation, namely the mean frequency differences between p-mode multiplets of odd and even degrees, which sample the Sun differently in latitude, 
and are therefore susceptible to 
solar activity;  Fossat et al. confined their analysis to 
p modes in the frequency range (2.32, 3.74) mHz.}, in the hope that there would be 
sufficient cancellation of near-surface activity for a g-mode signature to emerge from 
the noise.   To this end they considered the power 
spectrum of the power spectra of 8-hour segments of the GOLF signal, with start times 
separated by 4 hours (which we refer to as the g-mode cadence).    Any temporal variation in the location of a peak, which 
here we refer to as a p-mode peak, was regarded as an 
indicator of structural variation within the Sun.  
The data segments were chosen to be 8 hours long in order that they be short enough not to annihilate the signals from what was hoped to 
be many of the g modes, yet long enough to gain some idea of the p-mode frequency separations, the latter corresponding to a timescale of about 4 hours.  The resolution so obtained 
is insufficient to differentiate between the frequencies of different modes with like values of 
$n+\frac{1}{2}l$, let alone modes with like $n$ and $l$ but with different azimuthal orders.   However, the hope 
was that by analysing the whole 16.5-year GOLF time series the stable frequencies of the perturbing 
g modes would be revealed, and their rotational splitting measured.  To achieve that 
goal, first the power spectrum of the temporal variations in the location of the p-mode 
peaks in the power spectrum of the power spectrum of the GOLF p-mode signal was established, which it was hoped is a g-mode signal.  It could therefore reveal the g-mode frequencies, at least those associated with periods exceeding 8 hours. For such modes, the  uniformly spaced 
components of a rotationally split multiplet merge with the sequence of multiplet 
frequencies, which are uniformly spaced in period, 
so to disentangle them in a spectrum is not trivial.  Fossat et al. attempted to 
do so by computing the autocorrelation of that spectrum, with the intent of exposing the  
uniformly spaced component representing rotational splitting.  Here we call that 
the rotational diagnostic.

\citet{fossat_etal_2017} identified three peaks in their rotational diagnostic, 
near $\nu_1 = 210\,$nHz, 
$\nu_2 = 630\,$nHz and $\nu_3 = 1260\,$nHz.  For want of a better term, we call them the principal peaks.  Fossat et al. presumed that their frequencies  represent rotational splitting of dipole 
and quadrupole g modes according to equation (\ref{1.11}), and argued that those frequencies must be Doppler-shifted 
into the frame rotating with stationary combinations of the p modes, namely the frame 
rotating with cyclic frequency  $<\Omega>_{\rm p}/2\pi$ whose value they took to be  $433.5\,$nHz, being an average of $434.5\,$nHz reported by 
\citep{Kommetal_2003ApJ...586..650K} and $432.5\,$nHz reported by 
\citep{fossatetal_2003ESASP.517..139F}\footnote{
\citet{DOG_solarmagnet_2017SoPh..292...70G} has suggested that the temporal variation 
of the geomagnetic field maps the 
rotation of the Sun's radiative interior, revealing an angular velocity $434.6\,$nHz sidereal which, after multiplication by the mean Coriolis parameter $\overline C = 0.9951$ corresponding to the pertinent p modes, yields $<\Omega>_{\rm p}/2\pi = 432.5\,$nHz.}.  Thus, each frequency $\nu_k$ of the rotational diagnostic is given by 
\begin{equation}
2\pi\nu_k\simeq s(l,m):=|m(C_l<\Omega>_{\rm g}-<\Omega>_{\rm p})|\,,
\label{1.11a}
\end{equation}
where $C_l=1-1/[l(l+1)]$.  Taking into account that their procedure does not 
reveal the sign of the splitting, Fossat et al. found that the best fit to the data is obtained when 
the g-mode averaged angular velocity is 
$<\Omega>_{\rm g}/2\pi\,\, = 1277\pm 10$ nHz, granted that the uncertainty in  
$<\Omega>_{\rm p}/2\pi\,$ 
is  $\pm 10\,$nHz. From that fit, $\nu_1$ corresponds to the putative dipole-mode splitting, and 
$\nu_2$ and $\nu_3$ correspond to quadrupole-mode splitting for 
$m = 1$ and $m = 2$ respectively, the rms deviation of the fit being 2 nHz.  
Here, and henceforth, the index $m$ refers to the azimuthal order of pertinent g modes.
The angular velocity  $\Omega$ was assumed to be  
uniform both beneath and above some radius 
$r_{\rm c}$ in the Sun, with $\Omega = \Omega_{\rm c}$ where $r < r_{\rm c}$ and 
$\Omega=\Omega_0 := <\Omega>_{\rm p}\,$ where $r > r_{\rm c}$.  We shall assume likewise.  $<\Omega>_{\rm g}\,$ is, of course, 
an average over the entire inner radiative zone.  Finally, Fossat et al. assumed without comment 
that $r_{\rm c}/R  = 0.2$, where $R$ is the radius of the photosphere, and thereby 
inferred that the central angular velocity 
$\Omega_{\rm c}$ of the Sun exceeds the value in the surrounding radiative envelope by 
a factor 3.8.  It is interesting to note that the gravitational quadrupole 
moment that that implies is $J_2 \simeq 2.6 \times 10^{-7}$, some 30\% greater than 
the normally accepted value  \citep[cf.][]{2015SSRv_ISSI..196...15G}.

\begin{figure}[h!]
\centering
     \begin{center}
          \includegraphics[width=12.0cm]{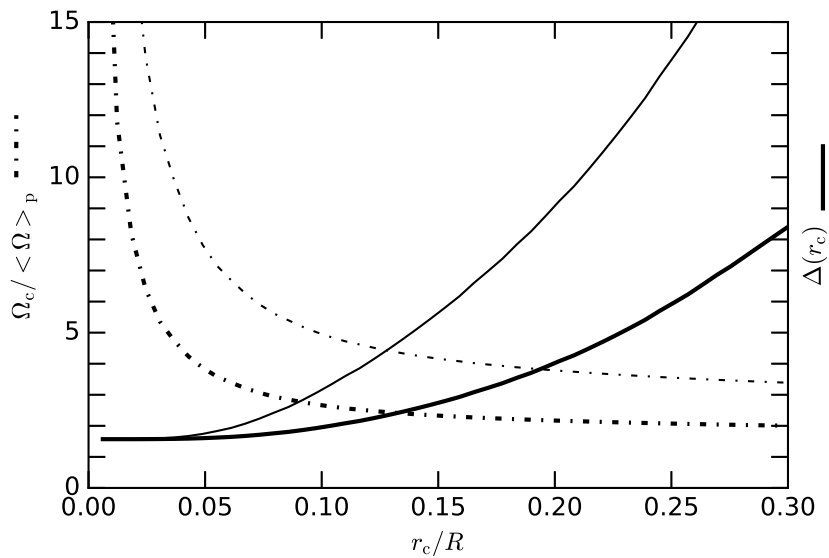}
     \end{center}
\caption{The dot-dashed curves are values of $\Omega_{\rm c}/<\Omega>_{\rm p}$ 
required to reproduce a g-mode rotational splitting $<\Omega>_{\rm g}$ 
from a presumed piecewise constant angular velocity 
$\Omega(r) = \Omega_{\rm c}\,$ where $r < r_{\rm c}$ and 
$\Omega(r) = <\Omega>_{\rm p}$  where $r > r_{\rm c}$,  with 
$<\Omega>_{\rm p}/2\pi = 433.5\,$nHz.  The thinner curves were computed with 
$<\Omega>_{\rm g}/2\pi\, = 1277\,$nHz, corresponding to assuming that 
g modes of degree 1 and 2 were detected by \citet{fossat_etal_2017}; the 
thicker curves correspond to assuming that
only the lowest even-degree g modes were detected, implying 
$<\Omega>_{\rm g}/2\pi\, = 789\,$nHz.  
The solid curves are the corresponding rms deviations $\Delta(r_{\rm c})$ from the GOLF p-mode rotational 
frequency splittings published by  \citet{golf_rot_split_garcia2004SoPh} and  
\citet{golf_rot_split_lazrek2004ESASP} 
of theoretical splittings resulting from the angular velocity $\Omega(r)$ in units of the standard errors in the splitting data. 
 A similar plot assuming $\Omega(r)$ to exceed $<\Omega>_{\rm p}$ by a half-Gaussian with $r_{\rm c}$ 
defined as the radius at which the excess is half its central value is hardly different.
\label{fig1}}
\end{figure}

One interesting property of this result is that the three putative g-mode sequences (each with varying $n$) that are present at apparently 
observable amplitudes  are associated
with values of $s(l,m)$ that form, presumably coincidentally, an harmonic sequence. 
Yet perhaps the most startling aspect of the result is that, at least in the form presented 
by \citet{fossat_etal_2017}, it is contradicted by inferences of the Sun's internal 
rotation obtained from direct measurements of rotational splitting of p modes.  For example, 
\citet{bison_rotation1995Natur} inferred from BiSON data that the core of the Sun appears 
to be rotating no faster than the rest of the radiative interior, and probably even more 
slowly.  And indeed, rotational splitting obtained from the GOLF data themselves 
\citep{golf_rot_split_garcia2004SoPh, golf_rot_split_lazrek2004ESASP}
are consistent with that inference.  
Moreover, the rms deviation of the p-mode rotational splitting implied by the piecewise  constant $\Omega$, with $r_{\rm c} = 0.2R$, is $\Delta = 8.9$ mean standard errors (in the rotational splitting data), whereas the deviation implied by uniform rotation is less than 1.6 standard errors. 
However, it cannot immediately be said that there 
is a genuine contradiction, because the p modes do not sense the very central regions of 
the Sun as much as do high-order g modes.  It should be appreciated that no constraint 
on $r_{\rm c}$ has been established; a small very rapidly rotating core cannot be negated by direct p-mode seismology.  The value of the core radius $r_{\rm c}$ 
adopted by \citet{fossat_etal_2017} is quite arbitrary, so it behoves one to consider 
other values.    In Figure \ref{fig1} we plot against $r_{\rm c}/R$ the rms deviation $\Delta$ from the published GOLF data of the p-mode frequency 
splittings implied by the assumed piecewise constant $\Omega$, in units of the rms uncertainties in the observations, 
choosing $\Omega_{\rm c}(r_{\rm c})$ to maintain the value of 
$<\Omega>_{\rm g}\,$ claimed  by \citet{fossat_etal_2017}.  $\Omega_{\rm c}/\Omega_0$ is also 
plotted. The theoretical splittings were evaluated using rotational splitting kernels 
computed from Model S of \citet{JCDmodelS1996Sci}.   A similar plot assuming $\Omega(r)$ to exceed $\Omega_0$ by a half-Gaussian with $r_{\rm c}$ 
defined as the radius at which the excess is half its central value: 
$\Omega(r)=\Omega_0 + (\Omega_{\rm c}-\Omega_0){\rm exp}[-({\rm ln}2)(r/r_{\rm c})^2]$ 
yields values for $\Omega_{\rm c}$ typically about 3\% greater. Taking the claim 
by Fossat et al. at face value, the mean g-mode frequency mismatch from equation (\ref{1.11a}) is about 4 standard errors.

It seems that the rapidly rotating core would need to be quite small if a significant  
inconsistency is to be avoided.  That would raise serious fluid dynamical issues, although 
it can be said that 
the shear layer between the two regimes, provided it were not too thin, could be dynamically 
stable at least to the Richardson criterion.  But there is another obvious, and rather more disturbing, issue that 
needs first to be addressed, namely that odd-degree g modes should not be detectable in 
the p-mode spectrum.  To that property we now turn our attention.

\section{On the strength of the g-mode coupling}
\label{sec:coupling}
The p-mode frequency perturbations can be estimated from equation (\ref{1.5}), using 
degenerate perturbation theory. 
The g-mode amplitude estimates reported in a second paper, by \citet{fossatschmider2018A&A},   imply that 
${|[\cal R}_{\boldsymbol{u}}+(\delta{\cal K}-\omega^2\delta{\cal I})/2\omega]/{\cal I}\omega|$
 is very small  (of order $10^{-5}$); therefore the deviation of the eigenfunctions from individual functions of the form 
given by equations (\ref{1.6}) and (\ref{1.7}) is also very small. 
\citet{kennedy_etal_1993} remarked that it had been shown  that the integral 
$\delta {\cal K} - \omega^2 \delta {\cal I}$
 vanishes for density and pressure perturbations 
with spherical-harmonic dependence of odd degree \citep{dog_L-H1993}, and concluded that g modes of odd degree 
cannot be detected from p-mode frequency changes.  Formally, that result had been derived for structure 
perturbations in hydrostatic balance.  However, it is straightforward to demonstrate that 
it holds also for any harmonic structure perturbation of odd degree, whether it be 
in hydrostatic balance or not, and also that the g-mode velocity 
contribution to the 
integral ${\cal R}_{\boldsymbol{u}}$ vanishes too (see Appendix~\ref{sec:appendixA}).  Therefore the assumption that dipole 
g modes are responsible for the principal 
peak near $\nu_1$ in the rotational diagnostic is thrown into very serious doubt.  One could, 
however,  adopt the assumption that it is quadrupole and hexadecapole modes that are 
responsible for the three principal peaks.  Applying the argument of \citet{fossat_etal_2017},  again recognizing the ambiguity of the signs of the splitting frequencies, then
leads to a slower core rotation, but still substantially faster than the surrounding  
radiative envelope.  
In that case, $<\Omega>_{\rm g}/2\pi\,\, = 789\pm 10$ nHz, 
and the rms deviation of the inferred splitting of the frequencies of ${\rm g}_{(l=2,m=1)} $ from $\nu_1$, ${\rm g}_{(l=4,m=2)}$ from $\nu_2$ and ${\rm g}_{(l=4,m=4)}$ from $\nu_3$, 
again when it is assumed that $r_{\rm c} = 0.2R$,  is 4.4 nHz, which is hardly different from the deviation according to the interpretation of Fossat et al.  The rms p-mode misfit from 
the GOLF frequencies \citep{golf_rot_split_garcia2004SoPh, golf_rot_split_lazrek2004ESASP} is then $\Delta =4.0$ standard errors.
The core rotation, and also the misfit to the p-mode rotational splitting 
frequencies, are plotted in Figure \ref{fig1} against putative core radii $r_{\rm c}$ up to $0.3R$.  

We add that, following \citet{fossat_etal_2017}, one might alternatively, and arguably more naturally, identify the 
largest  (210 nHz) principal peak in the autocorrelation with the detectable (prograde) g modes of 
lowest degree and azimuthal order, namely $l=2,\,m=1$;  that would imply a core rotating 
somewhat more slowly than the surrounding radiative interior, yielding  
$<\Omega_{\rm g}>/2\pi = 394\,{\rm nHz}$, which is consistent with earlier inferences from 
BiSON p-mode rotational splitting frequencies \citep{bison_rotation1995Natur,chaplin_etal_rotation_1999MNRAS.308..405C}.  However, 
it leaves the other principal peaks, if they be significant, unexplained.

Another outcome of the undetectability of  odd-degree g modes 
is that the argument by \citet{fossat_etal_2017} in their \S5 based on 
fitting a theoretical frequency distribution to the GOLF power spectrum  as a means of justifying the detection of 
dipole g modes must evidently be suspect. In \S \ref{sec:modelfitting} here we have more to report about theoretical model fitting for the purpose of establishing, or even  merely strengthening, a 
case for the existence of a proposed phenomenon.

Finally, we point out that in a second paper \citet{fossatschmider2018A&A} estimate, 
again by model fitting,  
some amplitudes of the components of the frequency variations of the p-mode peaks induced by what 
they believe to be the gravest (lowest order $|n|$) of the g modes that they studied.   
From that information one can make a crude estimate of the g-mode amplitudes from 
equation (\ref{1.5}), using expressions (\ref{A7}), (\ref{A8}) and (\ref{A9}) to
determine the coupling integrals.  A precise evaluation cannot be made, because 
we do not know the relative amplitudes of the many p modes to which the g modes were coupled.  If 
they were approximately equal, then the surface velocity amplitude of the quadrupole 
g mode $g_{36,l=2}$ would be of the order of 50 cm $\rm s^{-1}$; the 
uncertainty is comparable at least with the value.  Nevertheless, this estimate, 
being seriously in excess of the upper bound estimated by the Phoebus collaboration 
\citep{appourchauxetal_gmode_limit_2010A&ARv..18..197A}, sheds doubt on the fitting 
procedure.

\section{The form of the p-mode signature}
\label{sec:pmodesignature}

The low-degree p-mode spectrum is composed of contributions arising principally from the 
multiplet cyclic frequencies $\nu_{n,l}$ (where here $n$ denotes the order of the p mode) given approximately by equation (\ref{1.7a}), 
each of which, broadly speaking, is split by rotation (and further, but to a much 
lesser degree,  by the presence of 
g modes and any other non-spherical perturbation, which for the moment we ignore) into singlets separated by $<\Omega_{\rm p}>/2\pi$.  
The so-called small (multiplet) separation 
$d_{n,l} := \nu_{n,l}-\nu_{n-1,l+2} \simeq 2(2l+3)A_{\rm p}\nu_0^2/\nu_{n,l}$
cannot be resolved in a power spectrum of observations of 8 hr duration, and so each peak in the power spectrum is composed of a sum of modes of alternately odd and even degree. 
In view of the relative sensitivity of the whole-disc GOLF observations 
\citep[e.g.][]{jcddoglineshiftinterpretation1982MNRAS.198..141C}, the major contributors to each peak are alternately 
$l=0$ and $l=2$, and $l=1$ and $l=3$, whose frequencies are separated by about 10~$\mu$Hz and 17~$\mu$Hz respectively.  
Stochastic variations of the amplitudes of those contributing components are the major 
source of the frequency variation of the spectral peaks, whose magnitudes are comparable with the frequency separations of the 
components.  Those variations occur on timescales of typically a few days, the nominal lifetimes of the p modes. 
We illustrate in the Supporting Material (Figure \ref{SM4-1}) a typical 
resolved echelle spectrum in the frequency range adopted by \citet{fossat_etal_2017}, 
which displays, particularly at high frequency, the resulting spread of apparently 
resolved frequencies.  In view of the low sensitivity of whole-disc Doppler observations 
to modes with $l=3$, the frequencies of the odd-degree peaks in the power spectrum 
result almost entirely from the apparent frequency variation of the dipole p modes, which is much smaller 
than the frequency variation of the even-degree peaks, whose principal contribution comes from the variation of the relative surface velocity amplitudes of the more widely spaced (in frequency) radial and quadrupole  modes, producing frequency shifts of the p-mode peaks of   
amplitude about 4 $\mu$Hz.  (Note that radial p-mode frequencies are unaffected by the necessarily nonradial g modes.)  Superposed on these random frequency variations  
is a spectrum of phase-coherent g-mode-induced variations, which are of very much lower 
amplitude.  However, most of the g modes are expected to maintain phase over a period 
comparable with or greater than the 16.5-year duration of the GOLF observations, so there can be hope of detecting them. Did Fossat et al. succeed in doing that?

\section{Examination of the GOLF data analysis}
\label{sec:GOLFdataanalysis}

\begin{figure}
\plottwowide{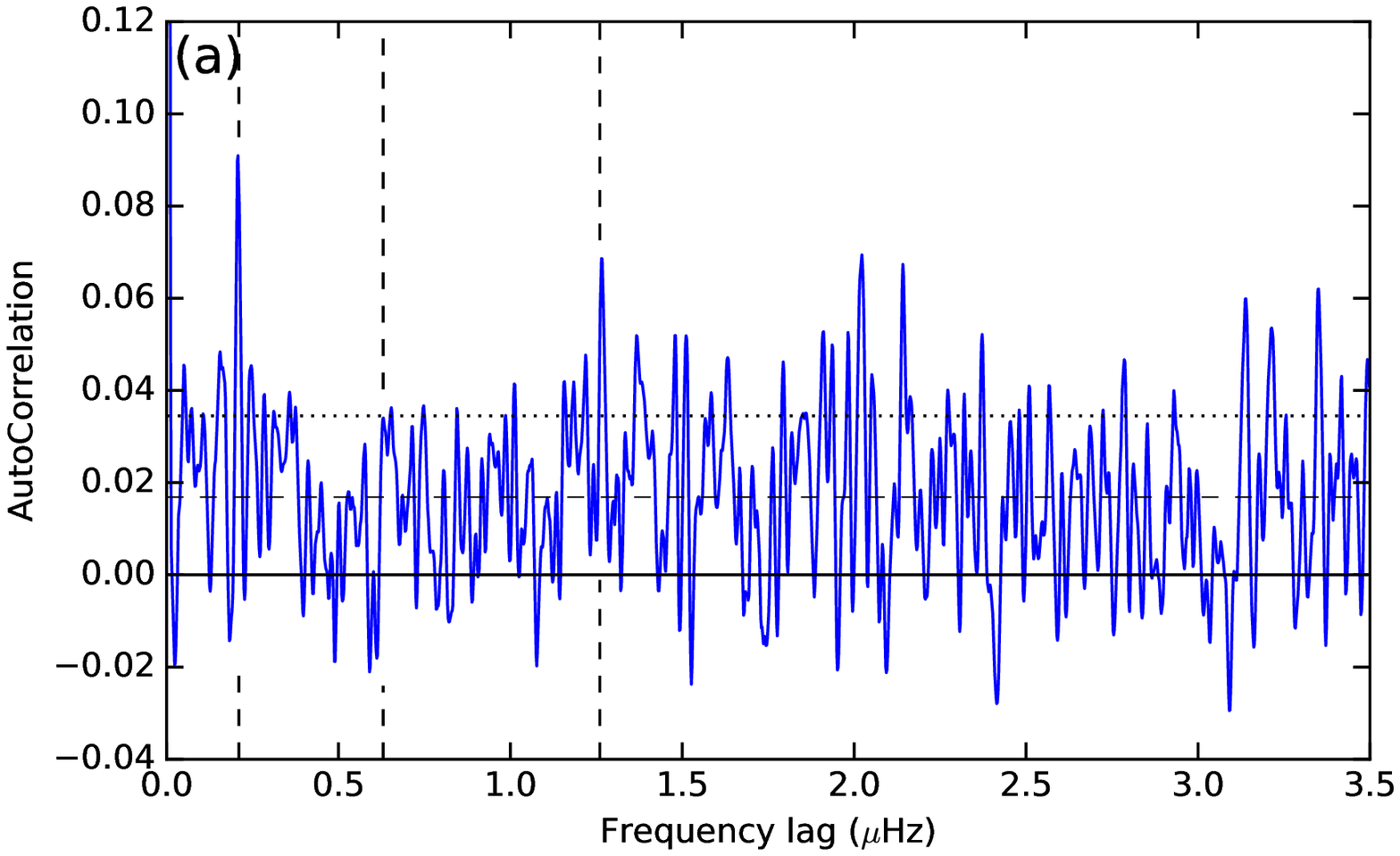}{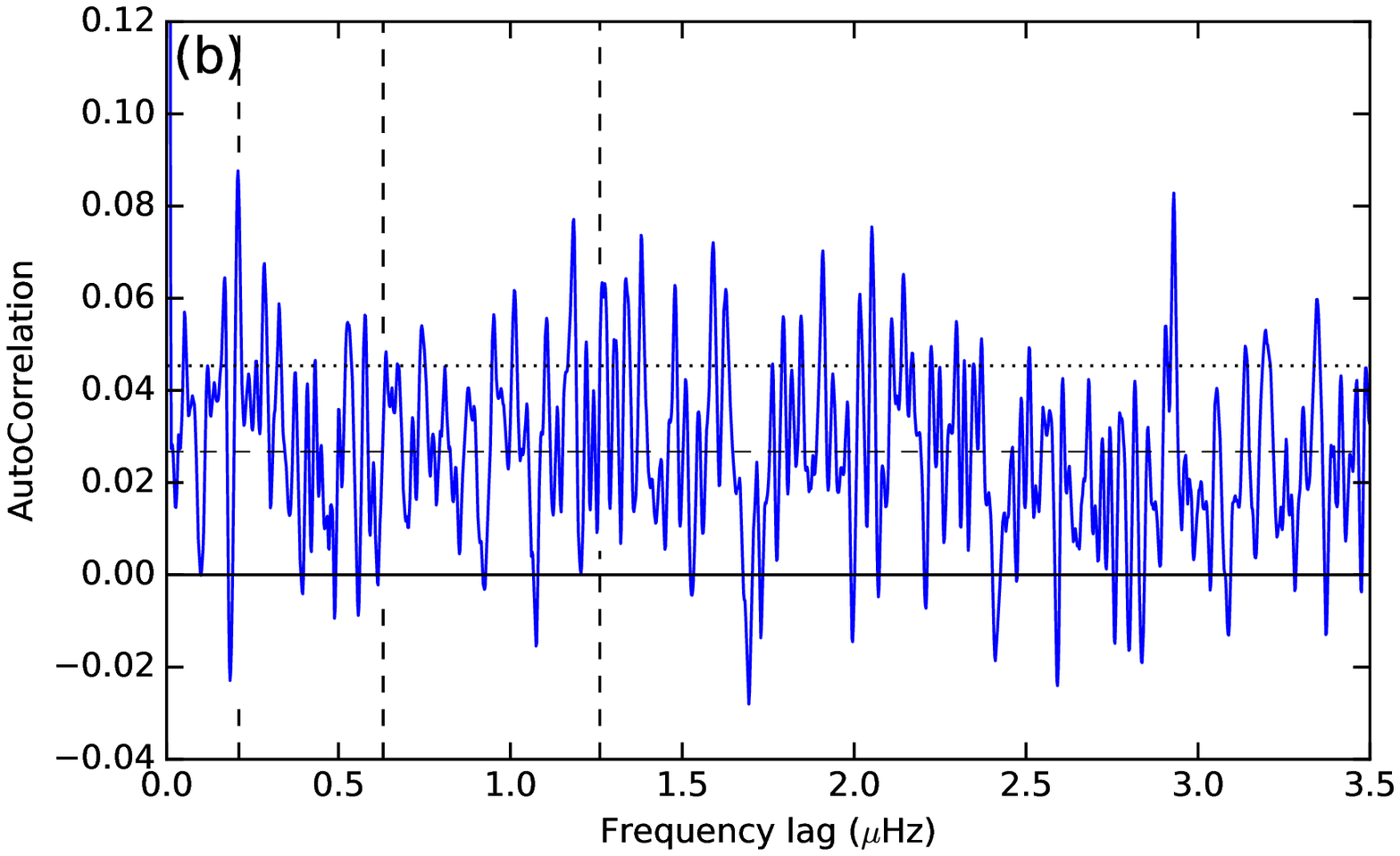}
\caption{Autocorrelation of power spectrum of p-mode peaks in power spectrum of power spectrum of GOLF data:  (a) from 80-second cadence,  (b) from 60-second cadence. The vertical dashed lines locate the frequencies 210 nHz, 630 nHz and 1260 nHz claimed by
\citet{fossat_etal_2017} to be g-mode rotational splitting frequencies. The horizontal solid, dashed, dotted reference lines show correlation levels 0.0, the mean from 0.023 to 3.5 $\mu$Hz, and one $\sigma$ above the mean respectively.
\label{fig2}}
\end{figure}

\citet{schunker_fragile_g-mode_detection2018SoPh..293...95S} have already essentially reproduced and 
extended some of the results reported by \citet{fossat_etal_2017}.
We first sought to do likewise, following Schunker et al. where the procedure was not 
described adequately by Fossat et al.   At our disposal were 
two 16.5-year  (1996-04-11 through 2012-10-05) whole-disc velocity observations obtained by the GOLF instrument, one averaged with 
a 60-second cadence, the other, which were the data of whose analysis Fossat et al. report, 
with a cadence of 80 seconds.  
We used 60 second GOLF data obtained from the SOHO data archive and 80 second cadence data provided by E. Fossat (personal communication, 2018)\footnote{The 60 second data are now available directly from the GOLF team at 
http://www.ias.u-psud.fr/golf/assets/data/GOLF\_velocity\_series\_mean\_pm1\_pm2.fits.gz and the 80 second data at https://www.ias.u-psud.fr/golf/assets/data/GOLF\_series\_Fossat\_et\_al.fits}
to match the same calibration used by \citet{fossat_etal_2017}. 
Each data set was divided into 8-hour overlapping segments whose start times were separated by 4 hours. Acceptable segments, namely those having duty cycles exceeding $90\%$, were zero-padded to $10^6$ seconds, and power spectra were computed. The 
segments of those spectra between 2.32 mHz and 3.74 mHz were extracted and divided by a Gaussian envelope function with standard deviation 0.39 mHz, centered at 3.22 mHz, and then 
 zero-padded to produce a frequency series in the range 0 -- 125 mHz. The spectrum of this series was then computed to produce a `period spectrum'. The dominant peak in this spectrum is usually near 14800 seconds. This period is associated with half the large p-mode separation, about $(67.54~ \mu{\rm Hz})^{-1} \simeq 14800~{\rm s}$. The period for each 8-hour segment was obtained by fitting to it an inverted parabola of total width 1600 s, as did Fossat et al.  This process
 produced a time series of 36132 possible values for the 80-second data and 36131 for 60-second data. In a two-step process first, ignoring times where insufficient data were available, the mean was removed, then times exceeding $\pm 240s$
 were removed, then the mean of the remaining 34095 useful representations of travel-time measurements for the 80-second data (34155 for the 60 second data) was removed and
 outliers and times with no data were set to zero.  The standard deviation for these two series is 52.0 and 51.1 for the 80 and 60 second series respectively. The power spectra of these 4-hour-cadence series were smoothed
  with a six-pixel-wide 
 boxcar, with weights (0.5,1,1,1,1,1,0.5), and the autocorrelation curves were finally obtained. 
The outcome from the two data sets are depicted in Figure \ref{fig2}, which is the analog of Figure 10 of Fossat et al.  The frequency lags $\nu_k$ = 210 nHz, 630 nHz and 1260 nHz, claimed 
to be g-mode rotational splitting frequencies, are indicated by the vertical dashed lines.   Panel (a) was obtained from the data with 80-second cadence; peaks at the frequencies $\nu_k$ are clearly evident, although the peak at $\nu_3$ is not as high as some of those 
that follow.   Panel (b) was obtained from the data with 60-second cadence; interestingly, 
not all the frequency lags $\nu_k$ correspond to prominent peaks.  The standard deviation 
$\sigma$ about the mean is the same, about 0.015, for both data sets.

It is disturbing that the two different averaging intervals (60 s and 80 s) of the same 
GOLF data set yield substantially different autocorrelation functions, particularly 
because the highest principal peak in one is hardly outstanding in the other.  We have repeated some 
of the tests carried out by \citet{schunker_fragile_g-mode_detection2018SoPh..293...95S}, 
such as the sensitivity of the results to: the cadence of the power spectra  of the GOLF data segments (4 hours in the case of the analysis by Fossat et al.);  the 
method of estimating the time interval of the 
p-mode peak in the power spectrum of those power spectra, which represents the inverse of 
half the large frequency spacing, approximately twice the sound travel time through a diameter of the solar interior;  the method of smoothing the power spectrum of 
those time intervals; and  ignoring a short segment of the GOLF time series, thereby 
introducing an offset in the segmentation of the data.  We are in broad agreement with the earlier work.

\section{MDI  and HMI data}
\label{sec:mdihmidata}

Even given the uncertainties described in the preceding sections, we considered it to be  worthwhile to use the 
same procedure from \citet{fossat_etal_2017}  as described above to examine other datasets with comparable sensitivity to p modes and comparable intervals of available data (see Supporting Material, Appendix~\ref{sec:appendixB}) for a list of data sources used.
We started with SOHO/MDI \citep{scherreretal1995SoPh..162..129S} data for the 15.5 year  interval 1996-05-01 through 2011.04.11 which is all but the final year examined by \citet{fossat_etal_2017}. 
MDI data processed for use in helioseismology studies are  available for spherical harmonic projections of degree 0 through 300.  The best match to the GOLF view of the Sun is to use the time series for degree 0, which is available as described in Supporting Material for section 6. 
Inspection of an echelle diagram similar to Figure \ref{SM4-1}, available in the Supporting Material as Figure \ref{SM6-1}, provides confirmation that MDI sees $l$ = 0 through 3 with relative sensitivity  similar to GOLF.   MDI observed solar photospheric motions at a 60-second cadence, and has nearly as complete a coverage as GOLF.  The MDI version of Figure 10  of \citet{fossat_etal_2017}, shown in Supporting Material as Figure \ref{SM6-2}, has a set of peaks superficially similar to those of the GOLF Figure 10, but none of them are near the three identified peaks seen in the GOLF figure.  It is interesting that the standard deviation of the semi-large-separation times for MDI is 58.8 seconds, close to the cadence of 60 seconds, for the available 30149 times in the span containing 31967 possible measurements.
 
We next examined the available 8.5-year interval from 2010-04-30 through 2018-10-20 from SDO/HMI (\citet{schouetal2012SoPh..275..229S}.  HMI is a higher-resolution version of MDI,  but with a cadence of 45 seconds and fewer data gaps.  We applied the same procedure as described above to analyse the observations for degree 0.  For HMI there are 17996 of a possible 18143 semi-large-separation measurements with the 4-hour cadence, the coverage   being 99\% compared to 94\% for MDI and GOLF, and the standard deviation of the set of times is less at 42.9 seconds.  There is a peak in the HMI version of Figure 10 near 1260 nHz (Figure \ref{SM6-3}), but it is only the fourth highest, and there are no substantial peaks at 210 nHz nor 630 nHz.  It is possible that effects of solar activity may have leaked into the analysis, or that the putative g-mode signal itself  varies with time.  We can compare the HMI result with that from GOLF by using data from the same interval.
 
 GOLF observations did not cease in 2012, and 
\citet{appourchauxetal_fossat_gmodes_2018A&A} have recently prepared a newly 
recalibrated 22-year GOLF dataset, now available directly from the GOLF project at \textit{https://www.ias.u-psud.fr/golf} for Level 2 data.
These GOLF data are available at a 20-second cadence,  allowing comparison of the 60-second and 80-second cadence versions of Figure 10 with the same calibration. 
Supporting Material Figure \ref{SM6-4} (a) and (b) show the results obtained in the interval 2010-04-30 through 2018-04-30.  The standard deviations of the times are 66.6 and 65.2 seconds for 80- and 60-second cadences respectively,  with 96\% coverage. 
\citet{fossat_etal_2017} found the peaks of interest in subsets of the GOLF data so one might have expected to see some similar 
pattern in the HMI data which shows a narrower distribution in the half-separation times than the GOLF data.
We have included "Figure 10" analyses of the \citet{appourchauxetal_fossat_gmodes_2018A&A} 80-second and 60-second data for the earlier 1996 to 2012 interval for comparison to Figure \ref{fig2} and \citet{fossat_etal_2017} Figure 10.  We see that the key 210 nHz peak is clear in this 80-second data as it was in the original data but not in the 60-second data.  In this case the same calibration of the GOLF data was used to generate the 80- and 60- second tests.

\section{Analysis of other observations}
\label{sec:otherobservations}

We have carried out similar analyses of other observations, namely by GONG (Global Oscillation Network Group), 
by the BiSON (Birmingham Solar Oscillations 
Network: \citet{2014MNRAS.441.3009D} and \citet{Hale2016} ), and by VIRGO/LOI (Variability of solar IRradiance and Gravity Oscillations Luminosity Oscillations Imager) 
on the SOHO spacecraft 
with broadly similar results (Figures \ref{SM6-6}, \ref{SM6-7}, and \ref{SM6-8} respectively.
In particular, the autocorrelations analogous to Figure 10 of \citet{fossat_etal_2017} and Figure 
\ref{fig2} here are superficially similar.  However, the principal peaks are not at the 
same frequency lags.  Evidently, the procedure is not robust.  Further details are 
presented in the Supporting Material.

\section{Offset start times}
\label{sec:offsets}

\begin{figure}[h!]
\centering
     \begin{center}
          \includegraphics[width=9.0cm]{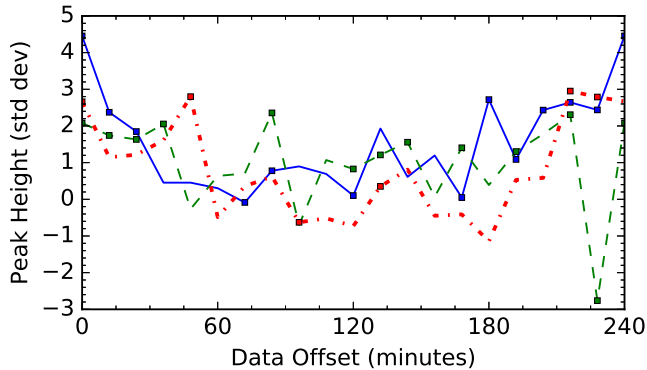}
     \end{center}
\caption{Heights of the purported g-mode peaks  with respect to the offset (in seconds) of the start time 
of the portion of GOLF data analysed.  Plotted are the greatest values of the autocorrelation, such as that in 
Figure \ref{fig2}  (but here in units of the standard deviation 
$\sigma$ in the entire frequency range 0 -- 3500 nHz), in a frequency window $\pm 10 $~nHz (the resolution width) about the frequencies $\nu_k$. They 
are joined by straight lines: continuous between peaks at about $\nu_1 = 210\, {\rm nHz}$, dashed about 
$\nu_2 = 630 \,{\rm nHz}$, and dot-dashed about $\nu_3 = 1260 \,{\rm nHz}$.  The symbols 
mark peaks in the autocorrelation;  where there is no symbol, there is no peak in the frequency range considered, the value plotted then being the autocorrelation at one of the limits of the window.
\label{fig3}}
\end{figure}

The effect of offsetting the start times of the data analysed perhaps provides some clue 
to interpreting the principal autocorrelation peaks evident in Figure \ref{fig2}.  In Figure 
\ref{fig3} we plot the heights of those peaks as a function of the offset.  They all 
drop sharply as the offset increases from zero, and remain low until the offset approaches the 4-hour g-mode cadence.  The autocorrelations at the offset of the g-mode cadence, and 
at any moderate integral multiple of it, are hardly distinguishable, because each 4-hour 
offset simply reduces the length of the 16.5-year data set by just one part in about 36000. 
However, a non-integral offset changes the phase of the data segment relative to terrestrial time, the effect of which is to reduce the autocorrelation.  So maybe the 
autocorrelation peaks result from perturbations in the spacecraft, such as voltage glitches 
caused by transmitting data to Earth, whose occurrences are linked to terrestrial time.  
We have searched for a 210 nHz frequency and its 
harmonics in those spacecraft procedures of which we are aware, and have found none.  Nevertheless, a terrestrially controlled  process on SoHO, either directly or indirectly 
related to GOLF, but not to MDI or VIRGO, remains a candidate for 
causing the autocorrelation peaks.  It is interesting that Fossat et al. chose their g-mode cadence to be 4 hours, which is an integral factor of a day; as \citet{schunker_fragile_g-mode_detection2018SoPh..293...95S} have pointed out, the principal autocorrelation 
peaks are smaller at different cadences.  That is the case  even for a cadence of 3 hours, which is also an integral factor of a day.  We have no explanation for that behaviour.

\section{Theoretical-model fitting}
\label{sec:modelfitting}

Fossat and his colleagues carried out two model-fitting procedures to justify 
their interpretation of their analysis of the GOLF data.  The first, by 
\citet{fossat_etal_2017}, was a two-dimensional fit with respect to uniform asymptotic 
multiplet dipole period spacing $\sqrt{2}P_0$ according to the leading term on the right-hand 
side of equation (\ref{1.7c}) and uniform rotational frequency spitting, according 
to equation (\ref{1.11}).  We note, in passing, that the dipole assumption 
is in conflict with the expectation that g modes of only even degree are accessible.  The other procedure, reported in the second paper \citep{fossatschmider2018A&A}, was a simpler, one-dimensional, degree-by-degree fit to the GOLF p-mode power spectrum of the 
approximate g-mode frequency formula (obtainable from equations (\ref{1.7c}) and (\ref{1.11})):
\begin{equation}
\nu_{n,l,m}\simeq \frac{L}{(|n|+l/2+\alpha_{\rm g})P_0}+\frac{m}{2\pi}(1-L^{-2})<\Omega_{\rm g}>\,, 
\label{1.18}
\end{equation} 
designed to detect rotational splitting of modes of degree $l=3$ and $l=4$. 
Here we address only the second procedure.

Once again, odd values of the degree $l$ were considered by Fossat and Schmider, which should be hardly detectable by p modes.  The fittings were accomplished by cross-correlating peaks at frequencies given by equation 
(\ref{1.18}) with the power spectrum of the GOLF p-mode peaks, adjusting not only the values of the 
parameters $P_0$, $\alpha_{\rm g}$, $<\Omega_{\rm g}>$, and the relative amplitudes of the modes, but also the frequency range adopted for the cross-correlation, in order to maximize the height of the zero-lag peak.  In the expectation that the g-mode frequency splitting is symmetrical with respect to azimuthal order $m$, the cross-correlation was symmetrized against zero lag to ease the optimization.  

Such a procedure can surely be used 
to fine tune a formula for a signal whose source is assured.  However, we doubt that 
it can be used to prove the existence of such a source.  To support our opinion, we 
have carried out a broadly similar analysis for a theoretical asymptotic spectrum of combined quadrupole and hexadecapole g modes in the frequency range 4--35 $\mu$Hz, selecting the relative amplitudes of the modes in such as way as to enhance both the zero lag and two `split' components at $\pm 210$~nHz.  An example of the symmetrized cross-correlation is depicted in Figure \ref{fig4}.  The frequency splitting in 
equation (\ref{1.18}) was not included: it was not necessary because each entry in the power spectrum against which the theoretical spectrum was cross-correlated was random, drawn from a Boltzmann distribution, and so contained no information whatever about rotational splitting, or even g-mode multiplet frequencies.  Moreover, we did not even adjust the other parameters in the right-hand side of equation (\ref{1.18}) to enhance the fit.  As in Figure 2 of 
\citet{fossatschmider2018A&A}, the  height of the peak at zero lag above the mean exceeds 10.5$\sigma$, where $\sigma$ is the standard deviation of the cross-correlation, and the height of the `splitting' peaks exceeds the 4.5$\sigma$.   To assess the significance of that result, we carried out 20 such analyses with different independently computed realizations of the Boltzmann distribution.  Of those, $10\%$ had a zero-lag height above 10.5$\sigma$, $45\%$ above 9$\sigma$ and $90\%$ above 8$\sigma$; the heights of the `splitting' peaks exceeded 
the 4.5$\sigma$  achieved by Fossat and Schmider in all cases.  We accept that our amplitude-adjustment procedure may have 
been different from that adopted by Fossat and Schmider (who did not report how their 
amplitudes were chosen), but we offer this exercise merely to warn against hasty inferences, 
not only that discussed here in support of the existence of rotationally split g-modes of degrees 3 and 4, but also the two-dimensional fitting reported earlier by \citet{fossat_etal_2017}.

Of course, from the point of view of this exercise the choice of a 210 nHz frequency lag for the cross-correlation of the 
theoretical spectrum with the random artificial power spectrum was arbitrary, and we 
could equally well have obtained similar results with a different lag.  Accordingly,  
we carried out the same exercise with the power spectrum of the GOLF data (obtained from the 80-s binning), attempting to reproduce peaks at 
lags other than frequencies $\nu_k$ of the principal peaks.  We found it to be 
significantly more difficult to achieve results as clean as those obtained from the 
random power spectra, such as that illustrated in Figure \ref{fig4}.  We therefore conclude 
that the frequencies of the p-mode peaks obtained from the GOLF data are not random.
However, we have no explanation for what the non-randomness might be.

\begin{figure}
\plottwowide{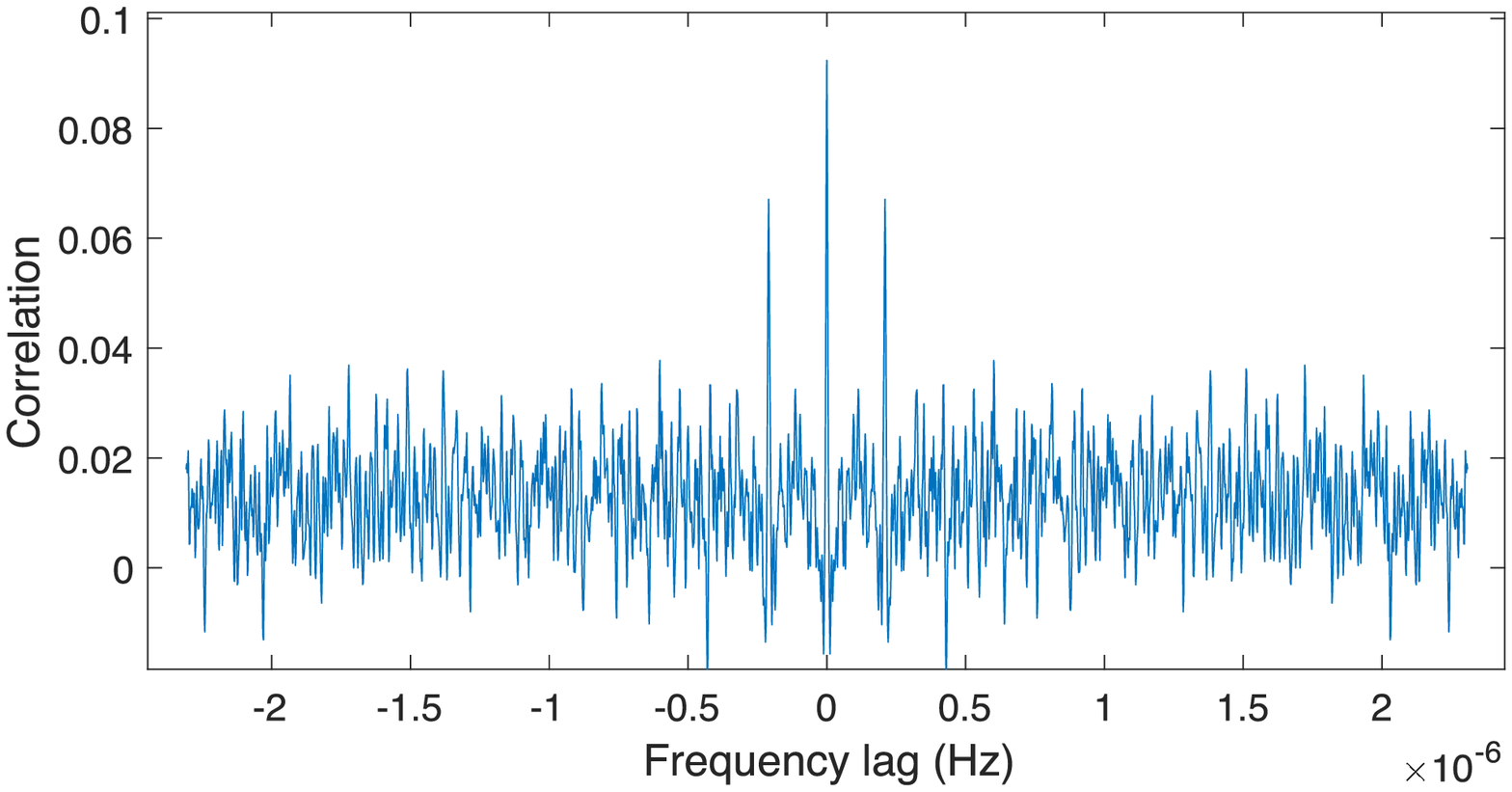}{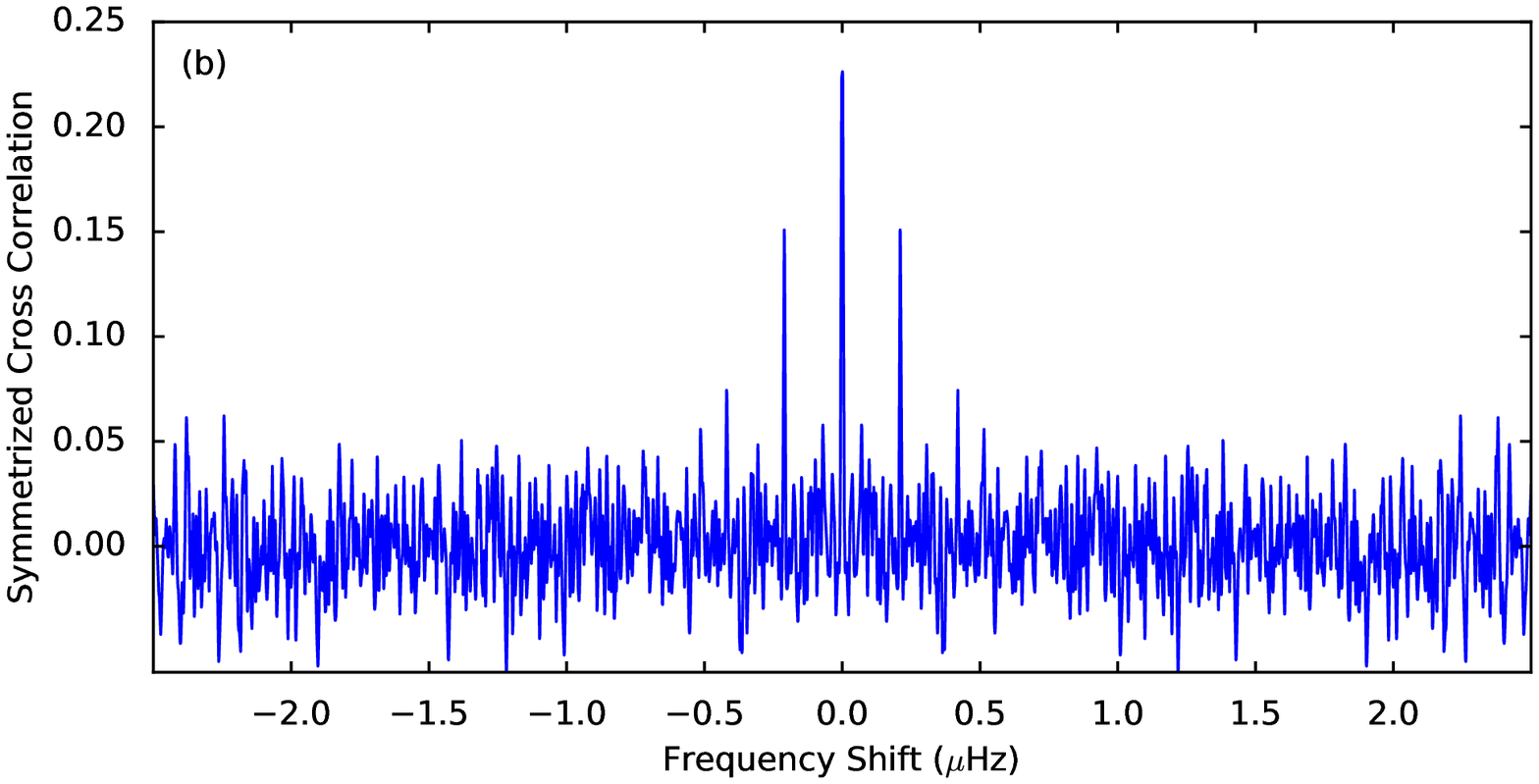}
\caption{Cross-correlation of an artificial g-mode power spectrum with a reference power spectrum:  
(a) Figure 1 of \citet{fossatschmider2018A&A}, 
whose reference is the large-separation spectrum of 80-second GOLF data,  
(b) the reference spectrum is random, each point having been drawn from a Boltzmann distribution.
(Part (a) reproduced with permission from Astronomy \& Astrophysics, \copyright ESO.) 
\label{fig4}}
\end{figure}

\section{Concluding remarks}
\label{sec:conclusion}

Our investigation has led us to doubt the report by \citet{fossat_etal_2017} of a 
detection of solar g modes via their interaction with p modes, and that a measurement 
of rotational splitting implies that the  core of the Sun is rotating rapidly.  The 
exact value of the angular velocity inferred at the centre of the Sun depends on an 
assumption of the variation with radius -- Fossat et al. assume that the region of rapid 
rotation extends to 20 per cent of the solar radius, from which they obtain a rotation rate 
4.8 times that of the surrounding envelope -- but that detail is not our major concern.   
Aside from the near inconsistency with direct inferences from
p-mode rotational splitting obtained from the same GOLF data by 
\citet{golf_rot_split_garcia2004SoPh} and \citet{golf_rot_split_lazrek2004ESASP}, and the 
reported g-mode amplitudes appearing to exceed the upper bound reported by the Phoebus 
group \citep{appourchauxetal_gmode_limit_2010A&ARv..18..197A}, the unconfirmed interpretation 
of the dominant modulation of the p-mode oscillations being due to dipole g modes is at 
odds with the property that odd-degree (static or slowly varying) spherically harmonic 
perturbations to the background state of the star do not modify p-mode frequencies in 
leading order.  Our neglect in the analysis we present explicitly here of the latitudinal variation 
of the angular velocity in the convection zone, and of the slow temporal variation of 
the g modes compared to the p modes 
\citep[e.g][]{lavelyritzwoller1992RSPTA.339..431L,hanasoge2017MNRAS.470.1404H}, makes no material difference to that conclusion. 
Fossat et al. appear to favour the assumption that the coupling, if it is detectable, 
is to g modes of lowest degree.  Therefore it is perhaps more natural in the 
first instance to adopt a more 
simplistic presumption:  that the dominant peak in their autocorrelation arises from quadrupole 
modes.  That does not explain the other peaks.  However,  we demonstrate here that those other 
peaks, and even the dominant peak, are not robustly determined.  Interestingly, the simplistic 
assumption implies that the Sun's core is rotating somewhat more slowly than the surrounding 
radiative envelope, which would be consistent with earlier findings from BiSON 
\citep{bison_rotation1995Natur}.   

Notwithstanding these immediate reactions, we have been led to investigate the analysis of 
the GOLF data more thoroughly, seeking to establish how robust are the conclusions of 
\citet{fossat_etal_2017} to modifications of the procedure adopted.  In agreement with a 
recent investigation by \citet{schunker_fragile_g-mode_detection2018SoPh..293...95S}, and 
noting that the record lengths and the cadence of the segmented data analysed by Fossat et al. 
are small factors of a terrestrial day, we have found that the dominance of the autocorrelation 
peaks depend critically on universal time.  We are also suspicious that the frequencies of 
the principal autocorrelation peaks, despite their apparent disparate physical origins, form 
an harmonic sequence.  Shifting the start times of the data segments 
by a small non-integral factor of a day reduces, or obliterates, the correlation, as does 
similarly changing the cadence.  Moreover, we find from analysing corresponding  seismic 
data from MDI, HMI, VIRGO, GONG and BiSON that although superficially similar autocorrelations 
emerge, the frequencies of the peaks do not coincide.  

We therefore surmise that the GOLF 
data have been influenced in some way by terrestrial processes that do not influence the other 
instruments in the same manner, and that the conclusions of Fossat et al. are premature.

Note added in revision: \citet{APPOURCHAUX.CORBARD2019} have performed similar tests and found similar results.

\section{Acknowledgments}
\label{sec:acknowledgments}
We are grateful to Todd Hoeksema and Thierry Appourchaux for constructive
discussion and Eric Fossat for providing the 80 second data he used.  We also
thank the SOHO GOLF and VIRGO/LOI teams for making their data available as per
the SOHO data access policy, and the Birmingham Solar-Oscillations Network
group for access to the BiSON data.
This work utilizes data obtained by the Global Oscillation Network
Group (GONG) program, managed by the National Solar Observatory, which
is operated by AURA, Inc. under a cooperative agreement with the
National Science Foundation. The data were acquired by instruments
operated by the Big Bear Solar Observatory, High Altitude Observatory,
Learmonth Solar Observatory, Udaipur Solar Observatory, Instituto de
Astrof\'{\i}sica de Canarias, and Cerro Tololo Interamerican
Observatory.
We acknowledge support by HMI NASA contract NAS5-02139.

\appendix

\section{g-mode velocity contribution}
\label{sec:appendixA}
As explained in the introduction, on the timescale of the diagnosing
p modes the temporal variation of the g modes can be ignored, permitting
instantaneous p-mode frequencies $\omega$ to be envisaged.  These are determined 
by equations (\ref{1.1}) - (\ref{1.4}) after linearization about the basic 
nonrotating equilibrium state of the star with respect to the velocity 
$\pmb v$ and the perturbed  structure $\Delta \rho$, $\Delta p$ resulting from the g modes.  The former is 
separated into pure (axisymmetric) rotation $\pmb \Omega$ and  
the (real) g-mode velocity ${\pmb u}$, and can be written 
\begin{equation} 
{\pmb v} =: {\pmb \Omega} \times {\pmb r} + \Sigma_{\rm order}\Sigma_l{\pmb u}_l\,,
\label{A1}
\end{equation}
where
\begin{equation} 
{\pmb u_l} = \Sigma_{m=-l}^l A_{l,m}{\pmb u_{l,m}} 
\label{A2}
\end{equation}
is the g-mode velocity component of degree $l$, with azimuthal component amplitudes $A_{l,m}$, in which 
\begin{equation} 
	2\,{\pmb u_{l,m}} = \left(W_l(r) P^m_l,\frac{U_l (r)}{L}\frac{{\rm d}P^m_l}{{\rm d}\theta},\frac{{\rm i}m U_l (r)}{L {\rm sin}\theta}P^m_l\right){\rm e}^{{\rm i}m\phi}\,+ {\rm c.c.}\,,
\label{A3}
\end{equation} 
$L=\sqrt(l(l+1))$ and c.c. denotes complex conjugate.  
For simplicity and without loss of generality, we take $W_l(r)$ and $U_l(r)$ to be real.
The outer sum in equation (\ref{A1}) is over all orders of the g modes; the 
amplitude functions $(W_l,U_l)$ are presumed to be normalized to unit 
inertia for each $l$.  To avoid unnecessary notational complication, we have omitted labels indicating mode order from $W_l\,,U_l$ and the amplitudes 
$A_{l,m}$.   The associated g-mode-induced pressure and density perturbation eigenfunctions, 
$\Delta p(r)  P^m_l({\rm cos}\theta)$ and $\Delta \rho(r) P^m_l({\rm cos}\theta)$, 
are related to the velocity eigenfunctions by the (adiabatic) oscillation equations 
\citep[e.g.][]{Unnoetal1989}:
\begin{equation} 
\frac{{\rm d}\Delta p}{{\rm d} r}+\frac{g}{c^2}\Delta p + (N^2-\omega_{\rm g}^2) \omega_{g}^{-1}\rho W=0,
\label{A4}
\end{equation}
\begin{equation} 
\Delta \rho = c^{-2}\Delta p + g^{-1}N^2\omega_{\rm g}^{-1}\rho W
\label{A5}
\end{equation}
in an obvious notation; $\omega_{\rm g}$ is the g-mode frequency.

The g-mode-induced perturbations to p-mode eigenfrequencies are given by equation 
(\ref{1.5}). Following normal degenerate perturbation theory, to leading order in 
the g-mode-induced perturbations, the resulting p modes of degree $\lambda$ can be represented as a sum over azimuthal order $\mu$ 
of all the zero-order eigenfunctions of degree $\lambda$ (and given order $n$) in the integrals in equation (\ref{1.5}):
\begin{equation} 
\pmb \xi_{n,\lambda,\alpha}=\Sigma_\mu c^\alpha_{n,\lambda,\mu}\pmb \xi_{n,\lambda,\mu}\,;\;\;\;\psi_{n,\lambda,\alpha}=\Sigma_\mu c^\alpha_{n,\lambda,\mu}\psi_{n,\lambda,\mu}\,,
\label{A6}
\end{equation}
where $\alpha$ labels the perturbed eigenmodes.  
The coefficients $c^\alpha_{n,\lambda,\mu}$ are determined by substituting this sum into equation (\ref{1.5}).  We do not present the entire analysis here, but merely inspect 
the contributions to the coupling integrals (\ref{1.2})-(\ref{1.4}) from a pair of p modes and a single g mode.  Beginning with the structure perturbations, we record that 
\begin{eqnarray}
& \delta_{\lambda,\mu,\mu',l,m}{\cal K} =  \int_{\cal V} \big( \gamma \chi'^* \chi \Delta_{l,m} p  
 + (\xi'^* \chi+\xi \chi'^*)\left[\frac{g}{c^2}\Delta_{l,m} p + 
(N^2-\omega_{\rm g}^2)\omega_{\rm g}^{-1}\rho W_{l,m}\right]  \nonumber \\
& + \frac{\rm d}{{\rm d}r}(g \xi'^* \xi)\Delta_{l,m}\rho \big)  P^m_l P^\mu_\lambda P^{\mu'}_\lambda{\rm e}^{{\rm i}(\mu-\mu'+m)\phi}{\rm d}V\, ,
\label{A7}
\end{eqnarray}
\begin{equation} 
 \delta_{\lambda,\mu,\mu',m}{\cal I} = \int_{\cal V} (\xi'^* \xi + \eta'^*\eta) \Delta_{l,m}\rho  P^m_l P^\mu_\lambda P^{\mu'}_\lambda{\rm e}^{{\rm i}(\mu-\mu'+m)\phi}{\rm d}V\,,
\label{A8}
\end{equation}
in an obvious notation, where
$\chi = {\rm div}\pmb \xi$ and $\gamma$ is the first adiabatic exponent; the asterisk denotes complex conjugate.    
The integrand has $P^m_l P^\mu_\lambda P^{\mu'}_\lambda {\rm exp}({\rm i}(\mu-\mu'+m)\phi]$ as a factor. Noting that 
there is another contribution $\delta_{\lambda,\mu,\mu',l,-m}{\cal K}$ to 
$\delta{\cal K}$, it follows that at least one of 
$\mu-\mu' \pm m$ must vanish for the sum of the two integrals not to vanish, from which it follows that $\mu-\mu'+m$ is even.  Note also that 
$P^\mu_\lambda$ is an odd or even function of ${\rm cos}\theta$ according to whether $\lambda+\mu$ is odd or even. Therefore 
$ P^m_l P^\mu_\lambda P^{\mu'}_\lambda$ is odd or even according to whether $l$ 
is odd or even, because $m+\mu+\mu'$ is even.  Hence the integrals vanish if 
 $l$ is odd.

Analysis of the advection integral ${\cal R}_{\boldsymbol{u}}$ is algebraically more complicated, but is otherwise essentially the same.  The component 
of the integrand analogous to the structure component (\ref{A7}), omitting subscripts to the oscillation eigenfunctions on the right-hand side,  is given by 
\begin{eqnarray}
& \pmb \xi^{*\mu'}_\lambda.({\pmb u_{l,m}}.\nabla) \pmb \xi^\mu_\lambda = 
\big(\big((W\frac{{\rm d}\xi}{{\rm d}r}-\frac{m \mu U(\xi -\Lambda^{-1}\eta)}{L r{\rm sin}^2\theta})\xi'^*  
+(\frac{\mu \mu'}{\Lambda^2 {\rm sin}^2\theta}W\frac{{\rm d}\eta}{{\rm d}r}+
\frac{m \mu'}{L \Lambda r {\rm sin}^2\theta}U(\xi-\frac{\mu^2}{L \Lambda {\rm sin}^2\theta}\eta))\eta'^*\big)P^m_l P^\mu_\lambda P^{\mu'}_\lambda  \nonumber \\
& +\frac{m {\rm cos}\theta \eta \eta'^*}{L \Lambda^2 r{\rm sin}^3\theta}UP^m_l (\mu P^\mu_\lambda \frac{{\rm d^2}P^{\mu}_\lambda}{{\rm d}\theta^2}+\mu' P^{\mu'}_\lambda \frac{{\rm d}P^\mu_\lambda}{{\rm d}\theta})   
+\frac{\mu \mu '\eta\eta'^*}{L \Lambda^2 r {\rm sin}^2\theta}U\frac{{\rm d}P^m_l}{{\rm d}\theta}(\frac{{\rm d}P^\mu_\lambda}{{\rm d}\theta}-{\rm cot}\theta\,P^\mu_\lambda)P^{\mu'}_\lambda   \nonumber \\
& +\Lambda^{-2}(W\frac{{\rm d}\eta}{{\rm d}r} - \frac{m \mu \eta U}{\Lambda r {\rm sin}^2\theta})\eta'^*P^m_l\frac{{\rm d}P^\mu_\lambda}{{\rm d}\theta}\frac{{\rm d}P^{\mu'}_\lambda}{{\rm d}\theta}  
+\frac{1}{L r}(\xi-\Lambda^{-1}\eta)\xi'^*U\frac{{\rm d}P^m_l}{{\rm d}\theta}\frac{{\rm d}P^\mu_\lambda}{{\rm d}\theta}P^{\mu'}_\lambda + \frac{\xi\eta'^*}{L \Lambda r}U\frac{{\rm d}P^m_l}{{\rm d}\theta}P^\mu_\lambda\frac{{\rm d}P^{\mu'}_\lambda}{{\rm d}\theta}  \nonumber \\
&+ \frac{\eta\eta'^*}{L \Lambda^2 r}U\frac{{\rm d}P^m_l}{{\rm d}\theta}\frac{{\rm d^2}P^\mu_\lambda}{{\rm d}\theta^2} \frac{{\rm d}P^{\mu'}_\lambda}{{\rm d}\theta}  \big){\rm e}^{{\rm i}(\mu-\mu'+m)\phi}\,,
\label{A9}
\end{eqnarray}
in which $\xi$ and $\eta$ represent components of the displacement eigenfunction of the p-mode of degree $\lambda$ (and order $n$) and azimuthal order $\mu$, and $\chi$ its divergence, and 
$\xi'$, $\eta'$ and $\chi'$ represent the corresponding mode of azimuthal order 
$\mu'$; also $\Lambda=\sqrt(\lambda(\lambda+1))$. Noting that the $\theta$ derivative of an odd function of ${\rm cos}\theta$ is even, and vice versa, it
is evident that all the terms in expression (\ref{A9}) have the same parity as $P^m_l P^\mu_\lambda P^{\mu'}_\lambda$.  
Consequently the argument in the previous paragraph applies here too.

More directly, expression (\ref {A9}) is real, which renders ${\cal R}_{\boldsymbol{u}}$ 
purely imaginary.   Therefore advection by the g-mode flow, to first order, does not 
influence the real part of the p-mode frequencies.  This comes about because, unlike 
rotational flow,  the essentially hydrostatic g-mode stream lines are all closed, and the 
local p-mode Doppler shifts cancel, as is the case also for meridional circulation 
\citep[e.g][]{DOGBWH2010ApJ...714..960G}.

Therefore, taking all the contributions to the p-mode frequency perturbations into account, it follows that
g modes of only even degree $l$ can influence the p-mode frequencies to leading order in 
the perturbations.

\section{Supporting Material}
\label{sec:appendixB}
The following pages will consitute the Supporting Material.

The figures are the backup for statements about the \citet{fossat_etal_2017}
Figure 10 processing applied to additional long duration helioseismic datasets
including GONG, VIRGO/LOI, BiSON. Also figures that support the use of SDO/HMI
data from the interval after the end of the Fossat et al. 2017 interval and
'echelle' format figures that support the use of l$l=0$ for MDI, LOI, HMI, and
GONG where the data is available in spherical harmonics versus data collected
from instruments that observer the Sun as a star with no spatial resolution
namely BiSON and GOLF.

The data sources used for the analyses shown are below.

\begin{enumerate}
\item GOLF 60 second data originally from the SOHO archive is now available from \url{http://www.ias.u-psud.fr/golf/assets/data/GOLF\_velocity\_series\_mean\_pm1\_pm2.fits.gz},
\item GOLF 80 second data used in  \citet{fossat_etal_2017} is now found at \url{https://www.ias.u-psud.fr/golf/assets/data/GOLF\_series\_Fossat\_et\_al.fits}\, data starts at 0:00:30 (T.A.I.) on April 11th 1996,
\item GOLF newer calibration 20 second data from \citet{appourchauxetal_fossat_gmodes_2018A&A} is now found at \url{https://www.ias.u-psud.fr/golf/assets/data/GOLF\_22y\_MEAN.fits},
\item MDI data is from the JSOC export tool at  \url{http://jsoc.stanford.edu/ajax/exportdata.html?ds=mdi.vw\_V\_sht\_gf\_72d[1996.05.01\_TAI-2011.02.12\_TAI][0][0][103680]&limit=none},
\item HMI data is from the JSOC export tool at \url{http://jsoc.stanford.edu/ajax/exportdata.html?ds=hmi.V\_sht\_gf\_72d[2010.04.30\_TAI-2018.10.21_TAI][0][0][138240]&limit=none},
\item GONG data is from \url{https://gong2.nso.edu/archive/patch.pl?menutype=t} for for $l=0$ for GONG months 1 through 225 useing start date 950507,
\item LOI $l=0$ data is from Thierry Appourchaux on 10 November 2017 and is available from him or Phil Scherrer on request,
\item BISON data is from \url{https://edata.bham.ac.uk/59/}, \dataset[doi:10.25500/eData.bham.00000059]{https://doi.org/10.25500/eData.bham.00000059}
\end{enumerate}

The software used to generate these figures can be found in the Stanford Digital Repository at \url{https://purl.stanford.edu/gt602xp9455}


\begin{figure}
\figurenum{B1}
\setcounter{figure}{4}
  \includegraphics[width=0.85\linewidth]{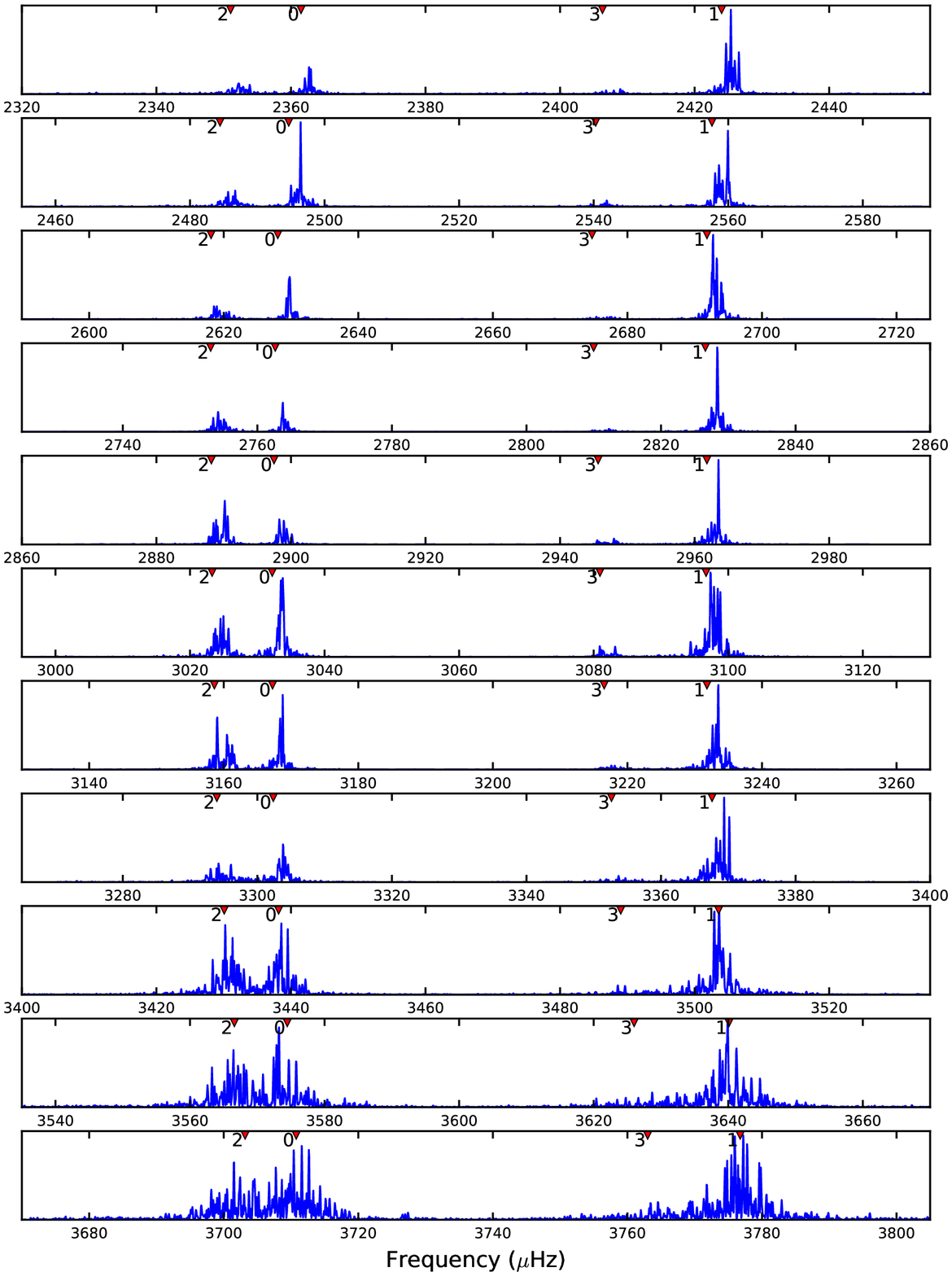}
  \caption{SOHO/GOLF echelle  diagram showing stacked sections of the GOLF
  p-mode spectrum after \citet{Grec_1983SoPh...82...55G}. Only the frequency
  range selected by \citet{fossat_etal_2017} is shown.  The data here is a 72
  day interval of the GOLF data starting in 1996.  To aid in identification the
  locations of the $l = 0,1,2,3$ p-mode frequencies from
  \citet{JCDmodelS1996Sci} model S are shown as red triangles at the top of
  each row with the $l$ value label.  Each row is scaled to the maximum power
  in that row.  The effect of shorter lifetimes with increasing frequency is
  evident. 
  \label{SM4-1}}
\end{figure}

\begin{figure}
\figurenum{B2}
\setcounter{figure}{5}
  \includegraphics[width=0.85\linewidth]{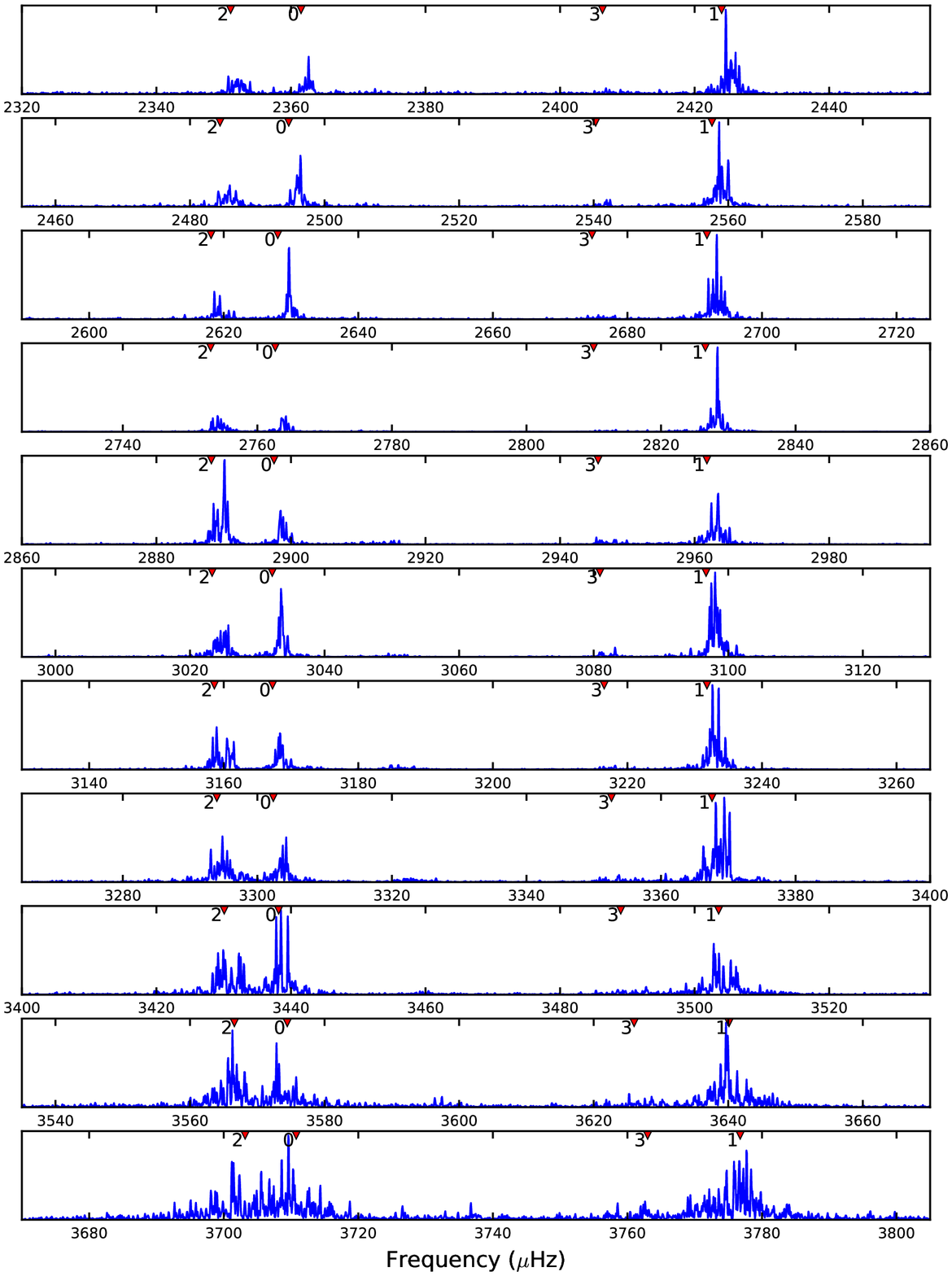}
  \caption{SOHO/MDI $l=0$ echelle diagram for the same interval as \ref{SM4-1}.
   The similarity of this figure and \ref{SM4-1} shows that using the $l=0$
  spherical harmonic for MDI and other instruments providing data as spherical
  harmonics  provides similar mode sampling as the unimaged GOLF data.
  \label{SM6-1}}
\end{figure}

\begin{figure}
\figurenum{B3}
\setcounter{figure}{6}
  \includegraphics[width=0.495\linewidth]{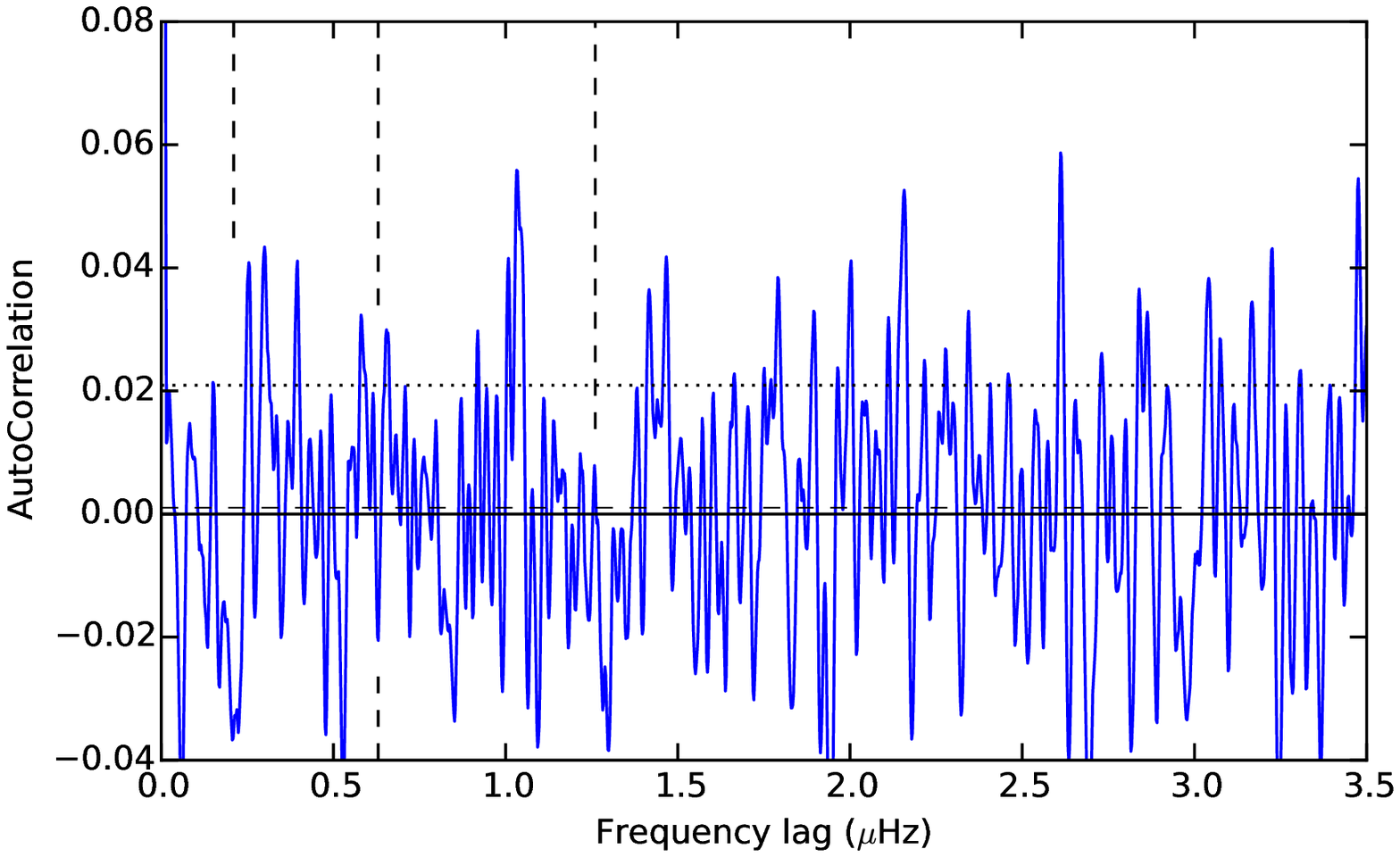}
  \caption{SOHO/MDI 60-second data for the 14.6-year interval from 1996-04-30
     through 2010-12-01 (\citet{scherreretal1995SoPh..162..129S})  processed the
     same as in Figure 10 of \citet{fossat_etal_2017} and Figure \ref{fig2} here.
     The vertical and horizontal lines are the same as in Figure \ref{fig2}.\label{SM6-2}}
\end{figure}

\begin{figure}
\figurenum{B4}
\setcounter{figure}{7}
  \includegraphics[width=0.495\linewidth]{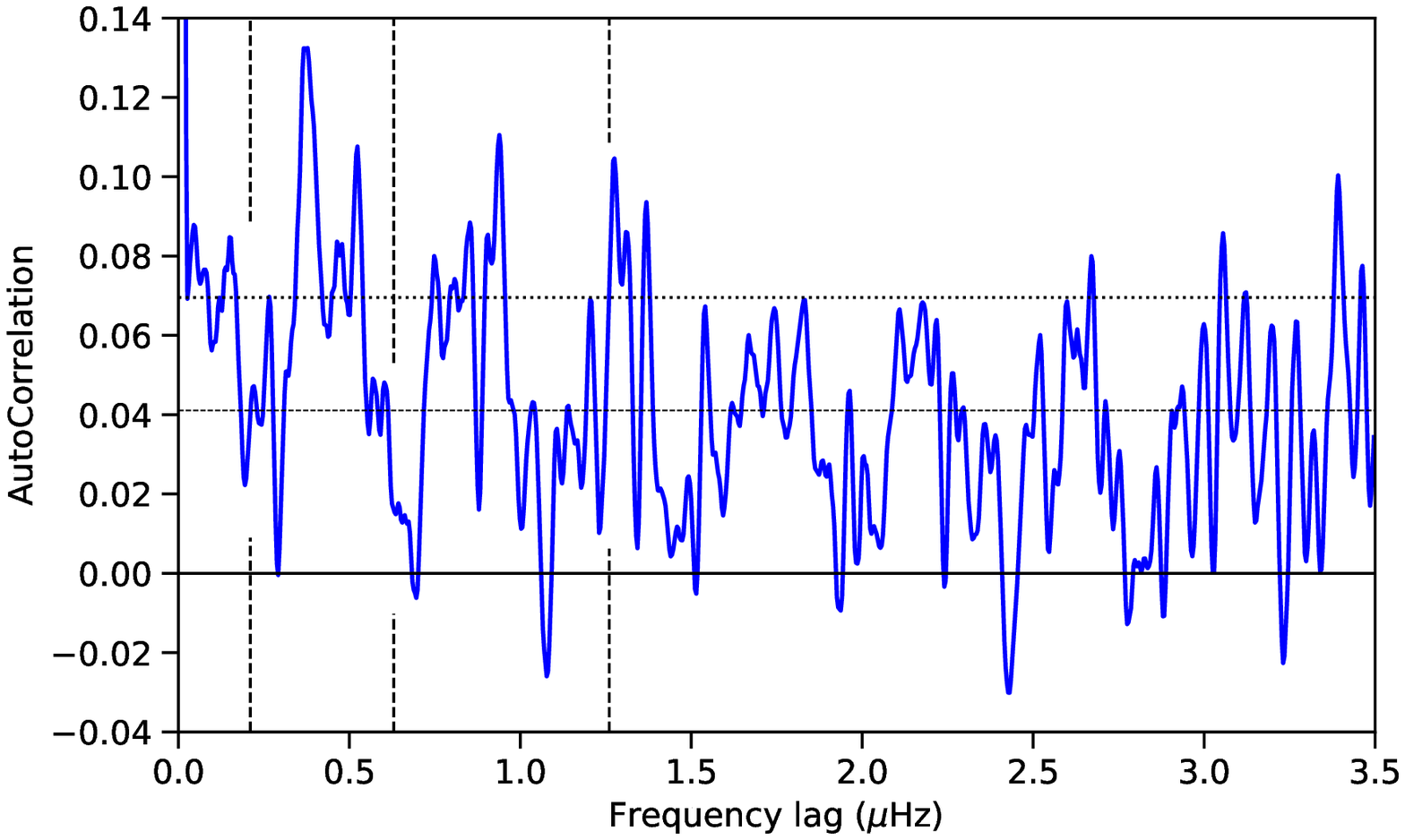}
\caption{SDO/HMI 45-second data for the  8.5-year interval from 2010-04-30
through 2018-10-20 (\citet{schouetal2012SoPh..275..229S})  processed the same
as in Figure 10 of \citet{fossat_etal_2017} and Figure \ref{fig2} here. \label{SM6-3}}
\end{figure}

\begin{figure}
\figurenum{B5}
\setcounter{figure}{8}
  \plottwowide{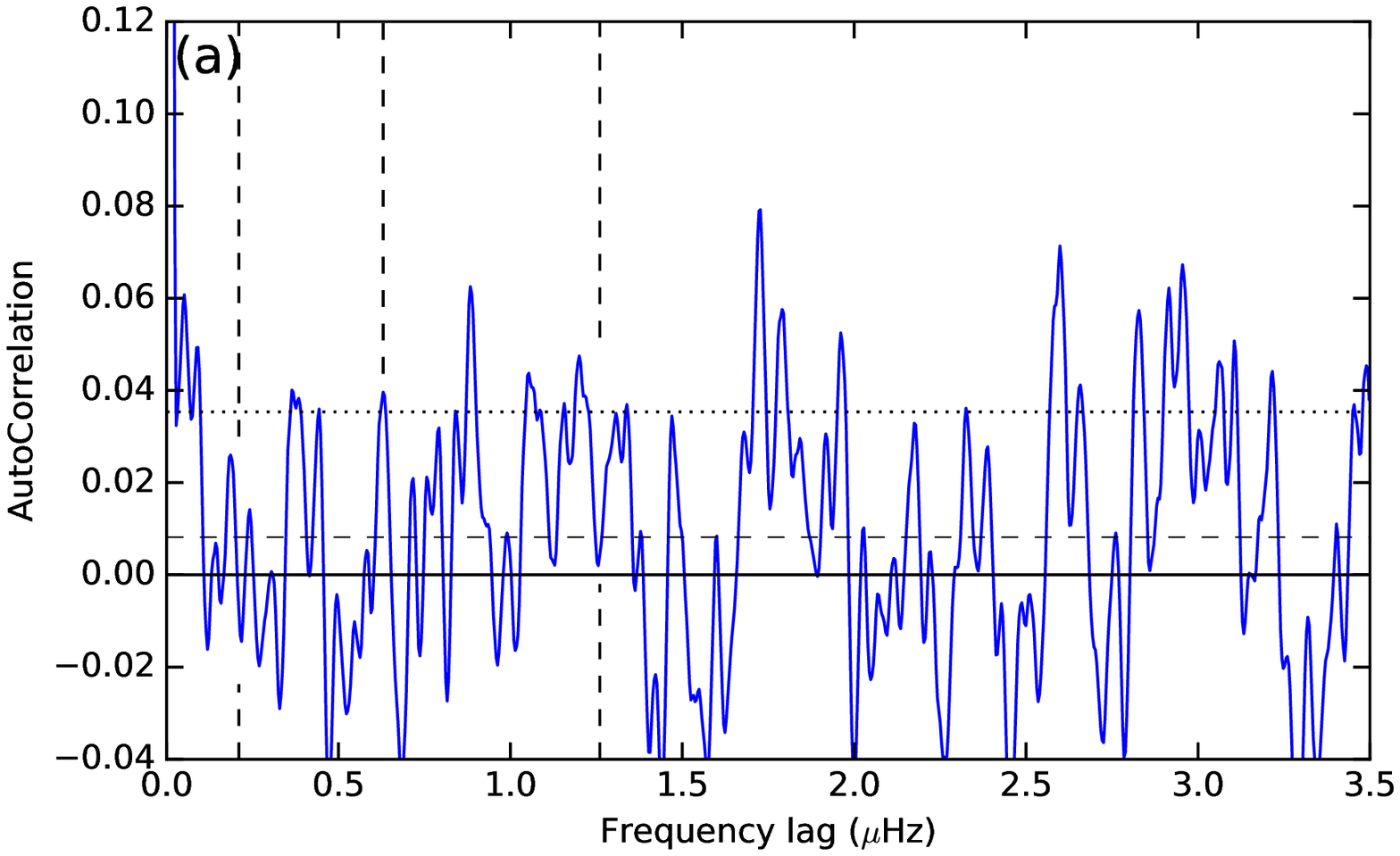}{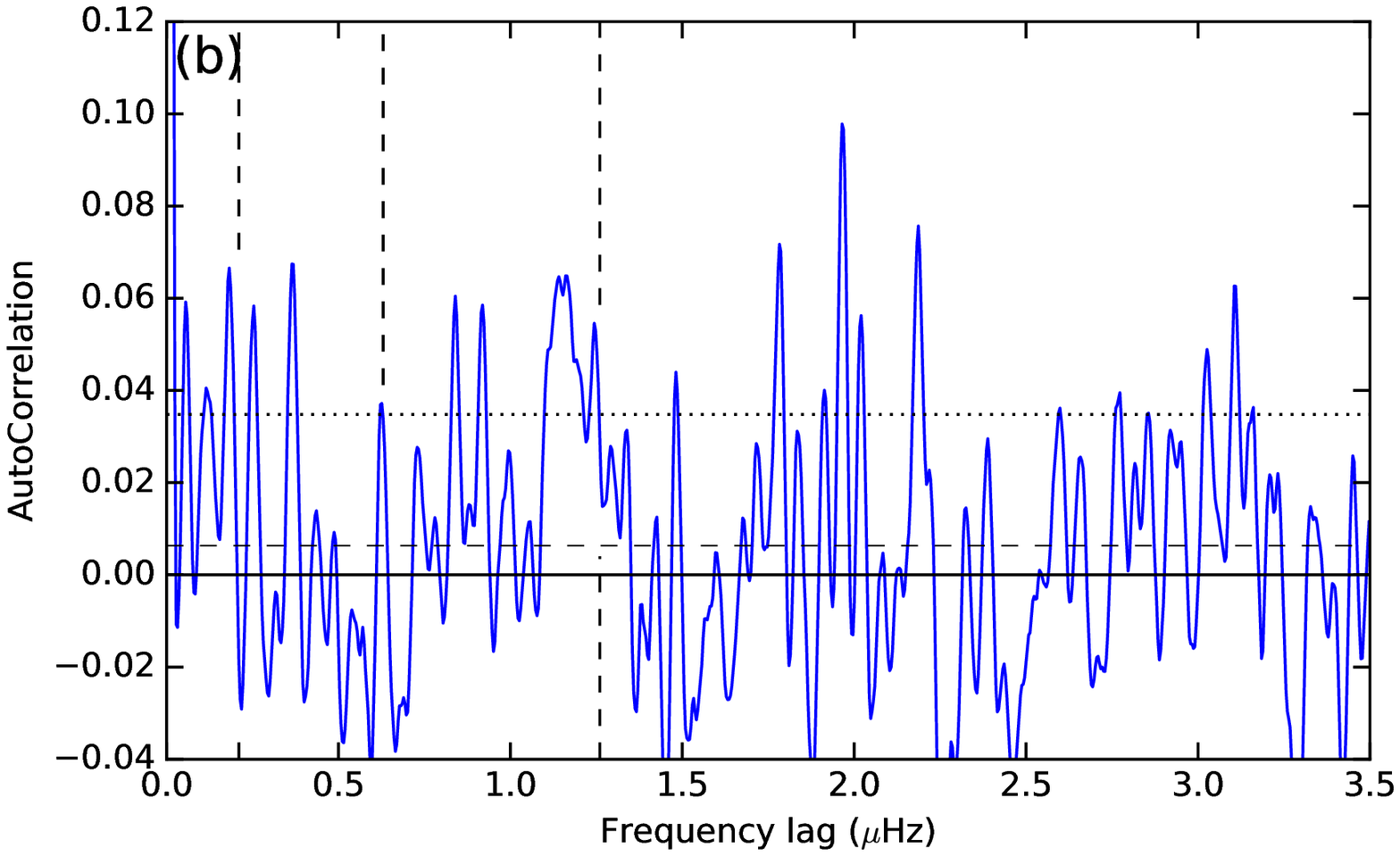}
  \caption{ GOLF data from \citet{appourchauxetal_fossat_gmodes_2018A&A} for
  the interval 2010.04-30 through 2018-04-30 for (a) 80-second and (b)
  60-second averages of the available 20-second data, processed the same as in
  Figure 10 of \citet{fossat_etal_2017} and Figure \ref{fig2} here for
  comparison to Figure \ref{SM6-3} HMI analysis. We note that the peaks seen in
  the 1996 to 2011 80-second data in both the \citet{fossat_etal_2017} Figure
  10 and  \citet{appourchauxetal_fossat_gmodes_2018A&A} calibrations as in
  Figure \ref{fig2} are absent in this later interval. 
  \label{SM6-4}}
\end{figure}

\begin{figure}
\figurenum{B6}
\setcounter{figure}{9}
  \plottwowide{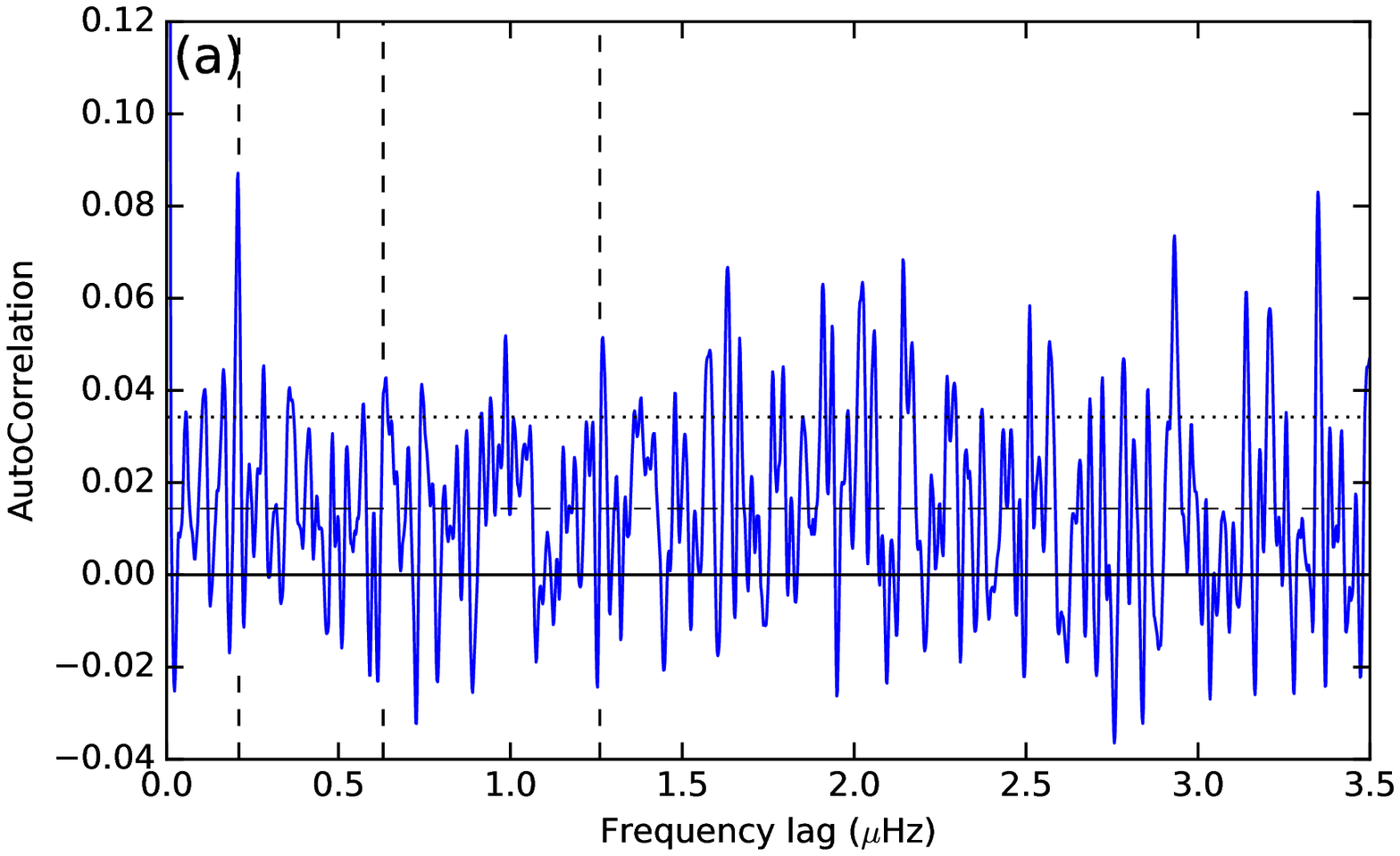}{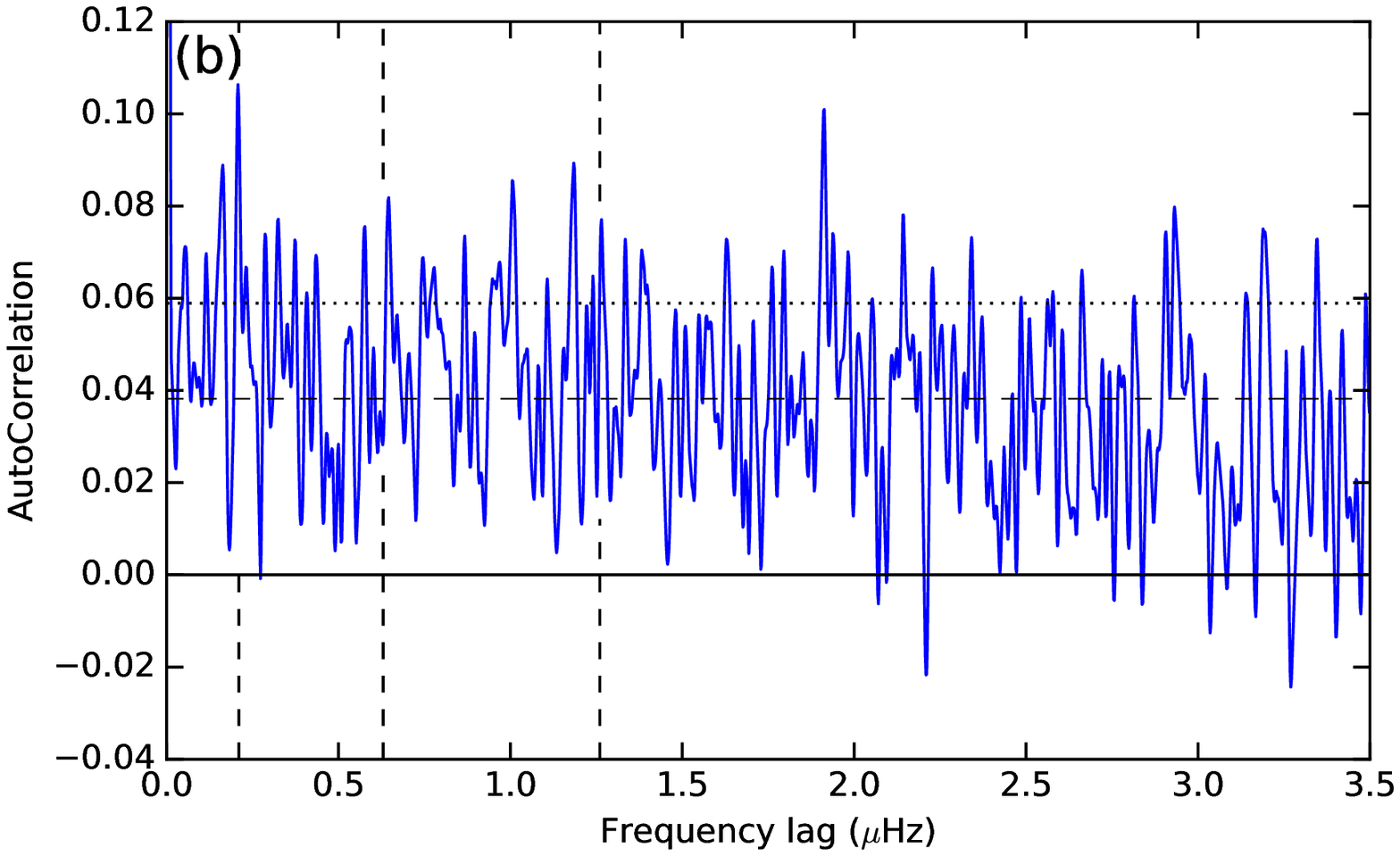}
  \caption{GOLF data using 80 and 60 second averges of the 20-second data from
   \citet{appourchauxetal_fossat_gmodes_2018A&A} in the same format as Figure
  \ref{SM6-4} but for the earlier 1996 to 2012 interval. As in Figure
  \ref{fig2} here which used data from an earlier GOLF calibration the 210 nHz
  peaks are somewhat more visible in the 80-second sampling. 
  \label{SM6-5}}
\end{figure}

\begin{figure}
\figurenum{B7}
\setcounter{figure}{10}
  \includegraphics[width=0.495\linewidth]{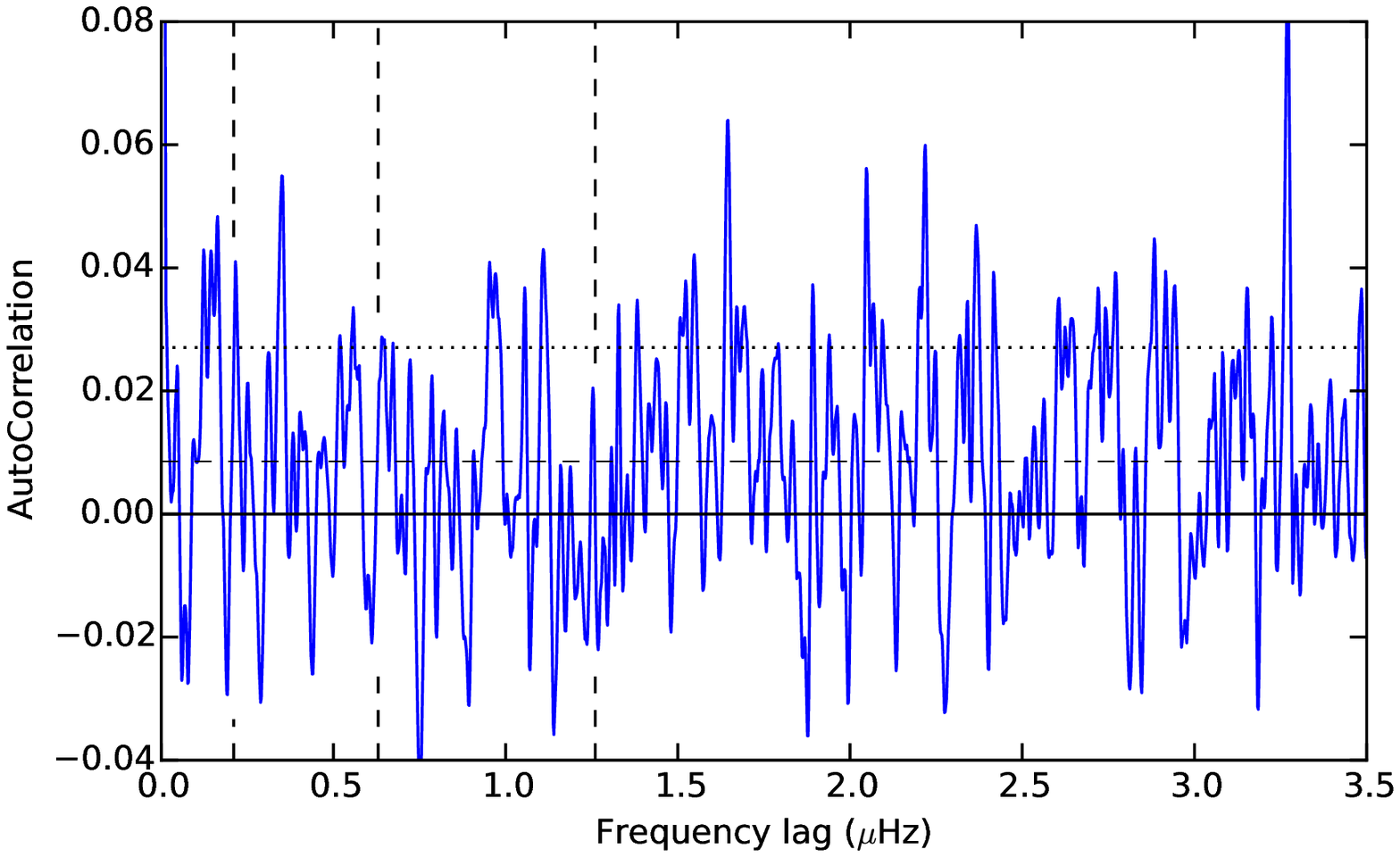}
  \caption{GONG 60-second data processed the same as in Figure 10 of
  \citet{fossat_etal_2017} and Figure \ref{fig2} here.  The 3 peaks at 210,
  630, and 1260 nHz are not significant. 
  \label{SM6-6}}
\end{figure}

\begin{figure}
\figurenum{B8}
\setcounter{figure}{11}
  \includegraphics[width=0.495\linewidth]{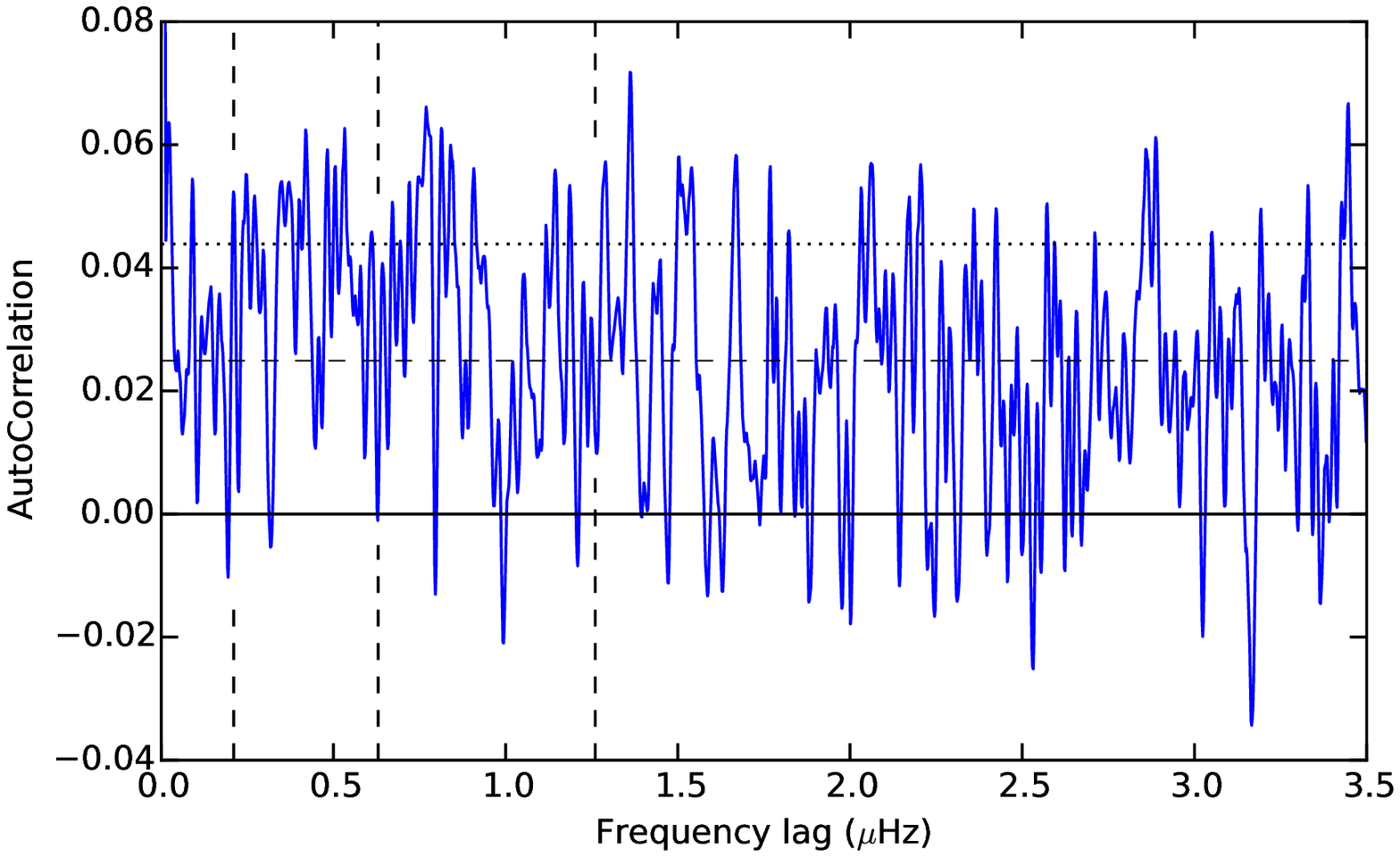}
  \caption{BiSON 40-second data processed the same as in Figure 10 of
  \citet{fossat_etal_2017} and Figure \ref{fig2} here.  The 3 peaks at 210,
  630, and 1260 nHz are not significant. 
  \label{SM6-7}}
\end{figure}

\begin{figure}
\figurenum{B9}
\setcounter{figure}{12}
  \includegraphics[width=0.495\linewidth]{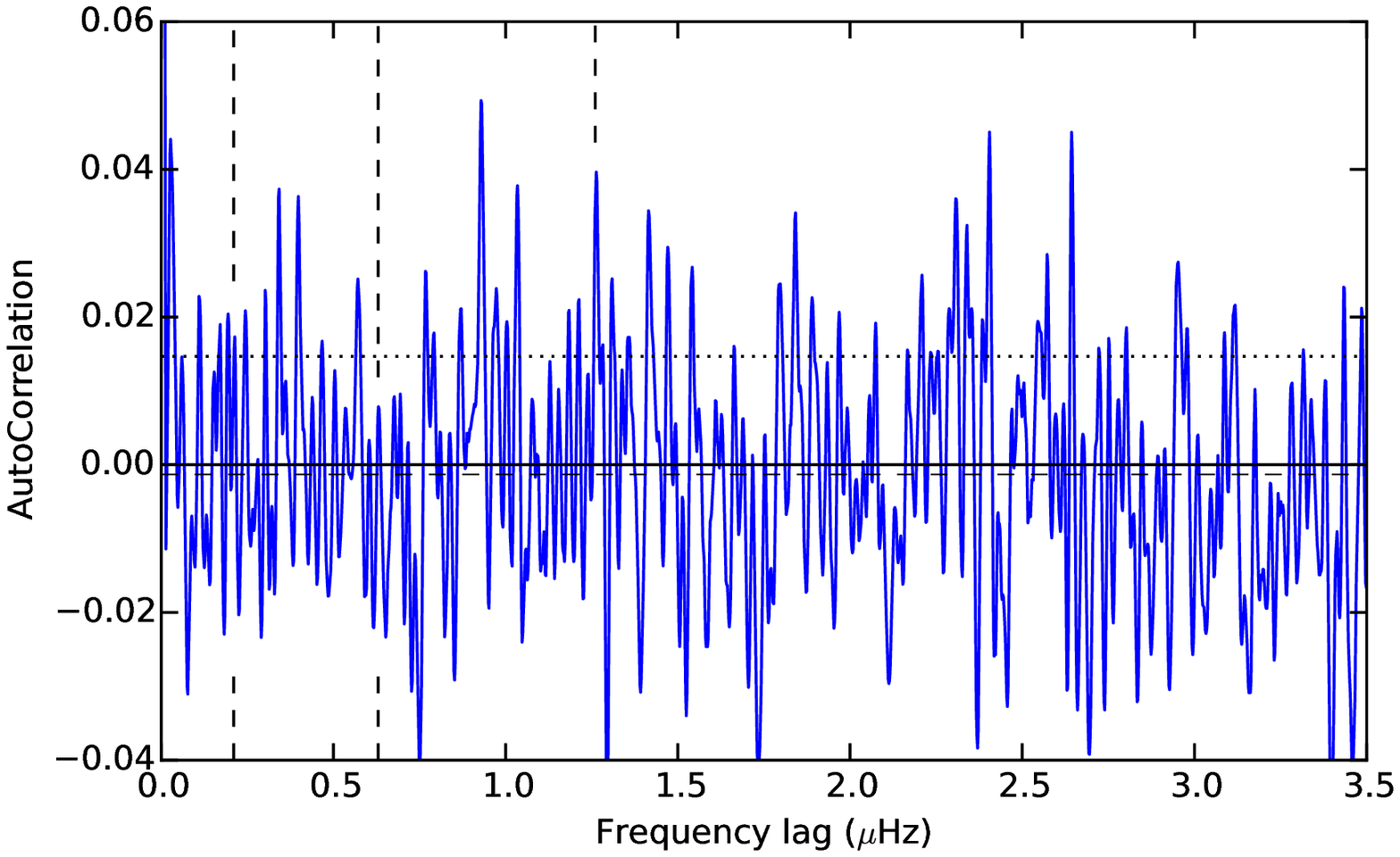}
  \caption{SOHO/VIRGO/LOI 60-second data processed the same as in Figure 10 of
  \citet{fossat_etal_2017} and Figure \ref{fig2} here.  The 3 peaks at 210,
  630, and 1260 nHz are not significant.
  \label{SM6-8}}
\end{figure}

\clearpage



\bibliographystyle{apj}
\bibliography{new.ms2}

%
%
%
%
%





\end{document}